\documentclass[12pt]{article}
\usepackage{epsfig,amsmath,amssymb,mathrsfs}
\usepackage[utf8]{inputenc}
\usepackage[colorlinks=true,citecolor=blue,,linktocpage=true,linkcolor=blue,urlcolor=black]{hyperref}
\usepackage{mathtools,slashed}
\usepackage{array}
\usepackage{xcolor}
\usepackage[force]{feynmp-auto}
\usepackage{caption}
\usepackage{subcaption}
\usepackage{overpic,youngtab}
\usepackage[latin,english]{babel}
\usepackage{amsmath}
\usepackage{amssymb}
\usepackage{epstopdf}
\usepackage{graphics,psfrag,rotating}
\usepackage{mathabx}
\usepackage{graphicx}
\usepackage{dcolumn}
\usepackage{float}
\usepackage{pdflscape}
\usepackage{array}
\usepackage{booktabs}
\usepackage{amscd}
\usepackage{mathtools}
\usepackage{fancybox}
\usepackage{fix-cm}
\usepackage[colorlinks=true,citecolor=blue,,linktocpage=true,linkcolor=blue,urlcolor=black]{hyperref}
\usepackage{tikz}
\usetikzlibrary{matrix}
\usetikzlibrary{positioning}
\usetikzlibrary{arrows}

  {\end{list}}%

\makeatletter
\renewcommand{\section}{\setcounter{equation}{0}\@startsection
 {section}%
 {1}%
 {0pt}%
 {-1\baselineskip}%
 {0.4\baselineskip}%
 {\bfseries\large}}%
\renewcommand{\subsection}{\@startsection
 {subsection}%
 {2}%
 {0pt}%
 {-0.75\baselineskip}%
 {0.2\baselineskip}%
 {\bfseries}}%
\renewcommand{\subsubsection}{\@startsection
 {subsubsection}%
 {3}%
 {0pt}%
 {-0.5\baselineskip}%
 {0.1\baselineskip}%
 {\sc}}%
\makeatother

\DeclareMathAlphabet{\mathpzc}{OT1}{pzc}{m}{it}

\usepackage{tikz}
\usetikzlibrary{patterns,snakes}
\usetikzlibrary{shapes.misc}
\usetikzlibrary{decorations.markings,decorations.pathmorphing}
\usetikzlibrary{calc}
\tikzstyle{spring}=[line width=0.8,black,snake=coil,segment amplitude=4.25,segment length=4.75,line cap=round]

\def\be{\begin{equation}}
\def\ee{\end{equation}}

\def\g5{\gamma_{5}}

\def\id3k{\int\!\! \dfrac{d^3\!\vec{k}}{(2\pi)^3 }}
\def\idthreeka{\int\!\! \dfrac{d^3\!\vec{k}_a}{(2\pi)^3 }}
\def\idthreek{\int\!\! \dfrac{d^3\!\vec{k}_3}{(2\pi)^3 }}
\def\idfourk{\int\!\! \dfrac{d^3\!\vec{k}_4}{(2\pi)^3 }}

\def\idthreex{\int d^{3}\!\vec{x}}

\def\idxtwo{\int d^{3}\!\vec{x}_2}

\newcommand{\bea}{\begin{eqnarray}}
\newcommand{\eea}{\end{eqnarray}}
\newcommand{\beann}{\begin{eqnarray*}}
\newcommand{\eeann}{\end{eqnarray*}}
\newcommand{\ba}{\begin{array}}
\newcommand{\ea}{\end{array}}

 \def\g {\gamma}

\newcommand{\email}[1]{\href{mailto:#1}{\tt #1}}

\begin{document}

\rightline{\scriptsize{FT/UCM 23-2025}}
\vglue 50pt

\begin{center}

{\LARGE \bf Tree-level  four-point function for a scalar field conformally coupled to Einstein's gravity on Euclidean $\text{AdS}_4$}\\
\vskip 1.0true cm
{\Large Carmelo P. Martin}
\\
\vskip .7cm
{
	
	{Universidad Complutense de Madrid (UCM), Departamento de Física Teórica and IPARCOS, Facultad de Ciencias Físicas, 28040 Madrid, Spain}
	
	\vskip .5cm
	\begin{minipage}[l]{.9\textwidth}
		\begin{center}
			\textit{E-mail:}
			\email{carmelop@fis.ucm.es}.
			
		\end{center}
	\end{minipage}
}
\end{center}
\thispagestyle{empty}

\begin{abstract}
Using the AdS/CFT correspondence, we compute the tree-level four-point boundary scalar correlation function for a scalar field conformally coupled to the graviton field on Euclidean $\text{AdS}_4$. We assume that the dynamics of the graviton field is governed by the Einstein-Hilbert action. We carry out the computation in momentum space and check that the result so obtained is consistent with the conformal Ward identities.
\end{abstract}

{\em Keywords:}  gauge/gravity duality, scattering amplitudes.
\vfill
\clearpage

\section{Introduction}

Since the early days of the analysis of the AdS/CFT correspondence, the computation of four-point correlation functions of the dual fields living on the boundary of Euclidean AdS (EAdS) has been of paramount importance \cite{Liu:1998ty, Freedman:1998bj, DHoker:1999kzh}: the dependence on the anharmonic or cross ratios of the four-point coordinates is not restricted by the conformal Ward identities. Over the years the study of the properties of the correlation functions of a conformal field theory has been carried out in coordinate space with the aim of fulfilling the conformal bootstrap program \cite{Poland:2018epd}. Indeed, the idea  is to take advantage of the expansion in conformal partial waves \cite{Dolan:2003hv} and the easy way --as compared to momentum space-- in which the conformal Ward identities can be solved \cite{Osborn:1993cr}.

And yet, some ten years  ago the analysis of conformal field theories in momentum space begun to be considered in earnest \cite{Bzowski:2013sza}. Since then, continuous advances have taken place in the area \cite{Bzowski:2015pba, Coriano:2019sth, Bzowski:2019kwd, Coriano:2019nkw, Bzowski:2020kfw}  --see  \cite{Coriano:2020ees}, for recent review. Alongside these studies the computation and discussion of the properties of the correlations functions of boundary fields for $\text{AdS}_{d+1}$ have been taking place --see \cite{Raju:2011mp}--\cite{Anero:2024bob}, for a partial list of references. For even space-time dimensions --remarkably, $d=3$--  the computations in question  are eased by the fact that the propagators in Poincar\'e coordinates involve Bessel functions with half integer index. To carry out such computations, thus obtaining explicit formulae, is relevant in order to get a better understanding of the gauge/gravity duality and, importantly, to  study the non-gaussian corrections to primordial fluctuations  generated during inflation, see \cite{Maldacena:2002vr}--\cite{Chowdhury:2023arc}.

The tree-level four-point boundary correlation function in momentum space for a conformally coupled scalar on an AdS background has been considered in a number of papers \cite{Oh:2020zvm, Chowdhury:2023arc, Albayrak:2020isk}. In these papers there is no graviton field. And yet, to the best of our knowledge, no such computation has been carried out when the conformally coupled scalar interacts with a graviton whose dynamics is governed by the Einstein-Hilbert action. In this paper we shall do this computation for Euclidean $\text{AdS}_4$. A nice feature of the theory that we shall consider is that it can be embedded into M-theory, as shown in \cite{deHaro:2006ymc}.

The layout of this paper is as follows. In section 2, we shall introduce the bulk theory which will be used to carry out the computations of the four-point boundary  correlation function for the field dual to the conformally coupled scalar field. In this section, the propagators needed to work out such correlation function will be given, along with the relevant lowest order perturbative contributions to the solution of the classical equation of motion of the graviton and scalar field. In section 3, we put the classical action on-shell as demanded by the gauge/gravity correspondence algorithm \cite{Witten:1998qj}, which relates the bulk theory and the boundary theory. The computation of the  four-point boundary scalar correlation function is carried out section 4. In section 5, we check that our result for the just mentioned boundary correlation function is consistent with the conformal Ward identities in momentum space. Closing comments are made in section 6.  Three appendices  are included, where details of the computations not displayed in the main text are given.

\section{A scalar field conformally coupled to Einstein's gravity on $\text{AdS}_4$.}

Our model will be that of a conformally coupled scalar field interacting with a graviton field on a Euclidean $\text{AdS}_{4}$ background. The dynamics of the graviton field
will be governed by the Einstein-Hilbert action. The complete classical action of our model reads thus:
\begin{equation}
\begin{array}{l}
 {S_{\text{\tiny{class}}}\,=\, S_{\text{HE}}+S_{\text{GHY}}+S_{B}+S_{\text{\tiny{scalar}}},}\\[8pt]
 {S_{\text{HE}}=\frac{2}{\kappa^2}\int_{\cal M}d^4 x\sqrt{g}\big(\,R[g_{\mu\nu}]-2\Lambda\big),}\\[8pt]
{S_{\text{GHY}}=\frac{4}{\kappa^2}\int_{\partial\cal M}d^{3} y\,\sqrt{g^{(b)}}K,}\\[8pt]
{S_{\text{B}}=-\cfrac{8}{\kappa^2}\int_{\partial\cal M}d^{3} y\,\sqrt{g^{(b)}},}\\[8pt]
{S_{\text{\tiny{scalar}}}=
 \frac{1}{2}\int_{\cal M}d^4 x\sqrt{g}\,\big(g^{\mu\nu}\partial_{\mu}\phi(x)\partial_\nu\phi(x)+\frac{1}{6}R[g_{\mu\nu}]\phi(x)\big)
 +\frac{\lambda}{4!}\int_{\cal M}d^4 x\sqrt{g}\,(\phi(x))^4.}
 \label{classaction}
 \end{array}
\end{equation}
In the previous equations, the symbol $\kappa$ is defined by $\kappa^2= 16\pi G$, $G$ being the Newton's constant, ${\cal M}$ stands for Euclidean $\text{AdS}_{4}$ and ${\partial\cal M}$ denotes its boundary. Euclidean $\text{AdS}_{4}$ ($\text{EAdS}_4$) is defined as the set of points $\{x=(z,\vec{x})\}$,
with $0<z<\infty$ and $\vec{x}\in \mathbb{R}^3$, where a Riemannian structure is defined by the line element
\begin{equation}
ds^2\,=\,\frac{1}{z^2}( dz^2+\delta_{ij}dx^i dx^j),\quad\quad i,j=1,2,3.
\label{EAdSline}
\end{equation}
Notice that we have set the Euclidean $\text{AdS}_4$ radius to 1. Also notice that, following \cite{Liu:1998bu}, we have included in $S_{\text{\tiny{class}}}$ the term $S_{\text{B}}$ to remove from $S_{\text{HE}}+S_{\text{GHY}}$ any contribution that is linear in the trace, with regard to the background metric, of the graviton field in the axial gauge, a gauge defined below. $(z,\vec{x})$ are called Poincar\'e coordinates.

Let us express the metric $g_{\mu\nu}(x)$ in (\ref{classaction}) as follows
\begin{equation*}
g_{\mu\nu}(x)=\bar{g}_{\mu\nu}(x)+\kappa h_{\mu\nu}(x),
\end{equation*}
where $\bar{g}_{\mu\nu}(x)$, the background metric, is the metric of $\text{EAdS}_4$ defined by (\ref{EAdSline}) and $h_{\mu\nu}(x)$ is the graviton field. Let us warn the reader that we shall work in the axial gauge
\begin{equation}
h_{z\mu}(x)=0,\quad \mu=z,1,2,3,
\label{axgauge}
\end{equation}
throughout this paper. $h_{ij}(z,\vec{x})$, $i,j=1,2,3$ will denote the, in general, non-vanishing entries of the graviton field. As usual Greek letter indices refer to indices in $\text{EAdS}_4$, whereas Latin indices refer to the 3-vector $\vec{x}$.

Let ${\cal O}(\vec{x})$ denote the scalar  operator on the $\text{EAdS}_4$ boundary dual to the scalar operator $\phi(x)$ in (\ref{classaction}). To obtain the tree-level contribution to the four-point function of  ${\cal O}(\vec{x})$ by using the AdS/CFT correspondence, we shall need the bulk-to-bulk propagators for $h_{\mu\nu}$ and the boundary-to-bulk and bulk-to-bulk propagators for the field $\phi(x)$.

In momentum space --we shall carry out our computations in momentum space, the bulk-to-bulk propagator for the graviton field $h_{\mu\nu}(x)$ in the axial gauge reads \cite{Raju:2011mp}
\begin{equation}
\begin{array}{l}
{
G_{i_1 j_1,i_2 j_2}(z_1,z_2;\vec{k})=(z_1 z_ 2)^{-1/2}\int_{0}^{\infty} d\omega\,\omega\, J_{3/2}[\omega z_1]\,{\tilde G}_{i_1 j_1,i_2 j_2}(\omega,\vec{k})\,J_{3/2}[\omega z_2],
}\\[8pt]
{
{\tilde G}_{i_1 j_1,i_2 j_2}(\omega,\vec{k})= G_1(T_{i_1 i_2} T_{j_1 j_2}+T_{i_1 j_2} T_{j_1 i_2}-T_{i_1 j_1} T_{i_2 j_2})+
}\\[8pt]
{
\phantom{G_{i_1 j_1,i_2 j_2}(\omega;\vec{k})=}
G_2(T_{i_1 i_2} L_{j_1 j_2}+L_{i_1 i_2} T_{j_1 j_2}+T_{i_1 j_2} L_{j_1 i_2}+L_{i_1 j_2} T_{j_1 i_2}  -T_{i_1 j_1} L_{i_2 j_2}-L_{i_1 j_1} T_{i_2 j_2}+
}\\[2pt]
{\phantom{G_{i_1 j_1,i_2 j_2}(\omega;\vec{k})=G_2(}
L_{i_1 i_2} L_{j_1 j_2}  +L_{i_1 j_2} L_{j_1 i_2}-L_{i_1 j_1} L_{i_2 j_2})+}\\[8pt]
{
\phantom{G_{i_1 j_1,i_2 j_2}(\omega;\vec{k})=}
 G_3(L_{i_1 i_2} L_{j_1 j_2}+L_{i_1 j_2} L_{j_1 i_2}-L_{i_1 j_1} L_{i_2 j_2}),
}
\label{BtBpropGR}
\end{array}
\end{equation}
where $J_{\nu}[u]$ stands for the Bessel functions of first kind and
\begin{equation}
\begin{array}{l}
{
T_{ij}=\vec{k}^2 \delta_{ij}-k_i k_j,\quad L_{ij}=k_i k_j},
\\[8pt]
{
G_1=-\cfrac{1}{2}\,\cfrac{1}{(\vec{k}^2)^2(\vec{k}^2+\omega^2)},\,
G_2=-\cfrac{1}{2}\,\cfrac{1}{(\vec{k}^2)^2\omega^2},\quad
G_3=-\cfrac{1}{2}\,\cfrac{1}{(\vec{k}^2) \omega^4}
}.
\label{coefBtBpropGR}
\end{array}
\end{equation}
We always use the definition $k=\sqrt{\vec{k}^2}$.
As it is customary \cite{Ammon:2015wua}, let us introduce an IR regulator in our bulk theory by moving the EAdS boundary from $z=0$ to $z=a$.
Then, the boundary-to-bulk \cite{Albayrak:2020isk}, $G(z,\vec{k})$, propagator and bulk-to-bulk \cite{Albayrak:2023kfk}, $G(z_1,z_2;\vec{k})$, propagators for $\phi(x)$ run  thus
\begin{equation}
\begin{array}{l}
{G(z,\vec{k})=\cfrac{\sqrt{\frac{2}{\pi}}\,k^{1/2} z^{3/2} K_{1/2}[k z]}{\sqrt{\frac{2}{\pi}}\,k^{1/2} a^{3/2} K_{1/2}[k a]},}\\[10pt]
{G(z_1,z_2;\vec{k})=(z_1 z_ 2)^{3/2}\int_{0}^{\infty} d\omega\,\omega\, J_{1/2}[\omega z_1]\,\cfrac{1}{\vec{k}^2+\omega^2}\,J_{1/2}[\omega z_2],}
\label{phiprops}
\end{array}
\end{equation}
respectively. The symbol $K_{\nu}[u]$ denotes the modified Bessel function of second kind.

In the AdS/CFT correspondence framework \cite{Ammon:2015wua}, to compute the tree-level contribution to the four point function of the operator ${\cal O}(\vec{x})$ one needs to solve first the Laplace equation --which is derived from the action in (\ref{classaction})--  for the field $\phi(x)$ on the EAdS background and for the appropriate boundary conditions for $\phi(x)$. The equation in question is
\begin{equation}
\bar{\nabla}_\mu\bar{\nabla}^\mu\phi(x)=\frac{1}{6}R(\bar{g})\phi(x),\quad  R(\bar{g})=-12.
\label{Laplaceeq}
\end{equation}
The bar above the covariant derivative shows that it is defined with regard to the background metric, $\bar{g}_{\mu\nu}$, defined by (\ref{EAdSline}).

Now, the conformal dimension, $\Delta$, of the dual ${\cal O}(\vec{x})$ \cite{Albayrak:2020isk} is given by $\Delta=d/2+\sqrt{d^2/4+1/6 R(\bar{g})}=2$, for $d=3$ and $R(\bar{g})=-12$. Hence, the Laplace equation in (\ref{Laplaceeq}) is to be solved \cite{Ammon:2015wua} with the boundary condition
\begin{equation}
\lim_{z\rightarrow a}\phi(z,\vec{k})= a^{d-\Delta}\phi_{0}(\vec{k})=a\, \phi_{0}(\vec{k}),
\label{bouncond}
\end{equation}
where $d=3$ and $\phi_{0}(\vec{k})$ is the Fourier transform of the source field for the operator ${\cal O}(\vec{x})$. Then, in momentum space, the solution, $\phi^{(0)}(z,\vec{k})$, to (\ref{Laplaceeq}), such that (\ref{bouncond}) holds, reads
\begin{equation}
\phi^{(0)}(z,\vec{k})= G(z,\vec{k})\, a\, \phi_{0}(\vec{k}).
\label{phizero}
\end{equation}
$G(z,\vec{k})$ is given in (\ref{phiprops}).

We shall also need the contribution, say $h^{(1)}_{ij}$, which is a part of the gravitational field solving the equations of motion derived from $S_{\text{\tiny{class}}}$  above and  which is of order $\kappa$ and it is quadratic in the scalar fields. This contribution is the following
\begin{equation}
h^{(1)}_{ij}(z_1,\vec{x}_1)=\kappa\int_{a}^{\infty}dz_2\idxtwo\,\sqrt{\bar{g}(z_2,\vec{x}_2)}\,G_{ij,mn}(z_1,z_2;\vec{x}_1-\vec{x}_2)\,{\cal T}_0^{mn}(z_2,\vec{x}_2),
\label{hache1ij}
\end{equation}
where
\begin{equation*}
\begin{array}{l}
{{\cal T}_0^{mn}(z,\vec{x})=\cfrac{1}{4}\,\bar{g}^{mn}\,(\phi^{(0)})^2+\cfrac{1}{6}\,\bar{g}^{mn}\phi^{(0)}\bar{\nabla}_\mu\bar{\nabla}^\mu\phi^{(0)}-
\cfrac{1}{12}\,\bar{g}^{mn}\bar{\nabla}_\mu\phi^{(0)}\bar{\nabla}^\mu\phi^{(0)}+}\\[4pt]
{\phantom{{\cal T}_0^{mn}(z,\vec{x})=}\,\cfrac{1}{3}\,\bar{\nabla}^m\phi^{(0)}\bar{\nabla}^n\phi^{(0)}
-\cfrac{1}{12}\,\phi^{(0)}(\bar{\nabla}^m\bar{\nabla}^n+\bar{\nabla}^n\bar{\nabla}^m)\phi^{(0)},}\\[10pt]
{G_{ij,mn}(z_1,z_2;\vec{x}_1-\vec{x}_2)=\id3k\,e^{i\vec{k}(\vec{x}_1-\vec{x}_2)}\,G_{ij,mn}(z_1,z_2;\vec{k}).}
\end{array}
\end{equation*}
In the previous equations, $\phi^{(0)}(z,\vec{x})$ and $G_{ij,mn}(z_1,z_2;\vec{k})$ are given in (\ref{phizero}) and (\ref{BtBpropGR}), respectively. The way $h^{(1)}_{ij}$ is obtained can be found is appendix A. An additional simplification can be reached by taking into account that $\bar{\nabla}_\mu\bar{\nabla}^\mu\phi^{(0)}(z,\vec{x})=-2\phi^{(0)}(z,\vec{x})$.

For future use, we next quote the following correction of order $\lambda$, say $\phi^{(1)}$, to $\phi^{(0)}$:
\begin{equation}
\phi^{(1)}(z,\vec{x})=-\cfrac{\lambda}{3!}\,\int_{a}^{\infty}dz_2\,\idxtwo\,\sqrt{\bar{g}(z_2,\vec{x}_2)}\, G(z_1,z_2;\vec{x}_1-\vec{x}_2)\,(\phi^{(0)}(z_2,\vec{x}_2))^3.
\label{phione}
\end{equation}
Where this $\phi^{(1)}(z,\vec{x})$ comes from is discussed in appendix A.

\subsection{ A simpler expression for $h^{(1)}_{ij}(z_1,\vec{x}_1)$.}

Let us introduce next the scalar field $\varphi^{(0)}(z,\vec{x})$ as follows
\begin{equation}
\begin{array}{l}
{\phi^{(0)}(z,\vec{x})=z\,\varphi^{(0)}(z,\vec{x}),}\\[8pt]
{\varphi^{(0)}(z,\vec{x})=\id3k\,e^{i\vec{k}\vec{x}}\,\varphi^{(0)}(z,\vec{k}),\quad
\varphi^{(0)}(z,\vec{k})=\cfrac{\sqrt{\frac{2}{\pi}}(k z)^{1/2}  K_{1/2}[k z]}{\sqrt{\frac{2}{\pi}}\,(k a)^{1/2}  K_{1/2}[k a]}\,\phi_{0}(\vec{k}).}
\label{varphi}
\end{array}
\end{equation}
Recall that $\phi^{(0)}(z,\vec{x})$ is defined by its Fourier transform, which is given in (\ref{phizero}).

After some computations, the substitution in (\ref{hache1ij}) of  the first equation in (\ref{varphi}) yields
\begin{equation}
\sqrt{\bar{g}(z,\vec{x})}\,{\cal T}_0^{mn}(z,\vec{x})=z^2\,{\cal V}_{0mn}(z,\vec{x}),
\label{TtoV}
\end{equation}
where
\begin{equation}
{\cal V}_{0mn}(z,\vec{x})=-
\cfrac{1}{12}\,\delta_{mn}\,(\partial_z\varphi^{(0)}\partial_z\varphi^{(0)}+\partial_i\varphi^{(0)}\partial_{i}\varphi^{(0)})+
\,\cfrac{1}{3}\,\partial_m\varphi^{(0)}\partial_{n}\varphi^{(0)}-\cfrac{1}{6}\,\varphi^{(0)}\partial_m\partial_{n}\varphi^{(0)}.
\label{simplervertex}
\end{equation}
In the previous expression $\partial_m$ denotes the partial derivative with regard to $\vec{x}$ --recall that Latin indices run from 1 to 3 and repeated indices signals sum over all of their values. To obtain (\ref{simplervertex}), we have used the equation $(\partial_z^2-\partial_i\partial_i)\varphi^{(0)}(z,\vec{x})=0$.

Next, the substitution of (\ref{TtoV}) in (\ref{hache1ij}) leads to
\begin{equation}
h^{(1)}_{ij}(z_1,x_1)=\kappa\int_{a}^{\infty}dz_2\idxtwo\,\,G_{ij,mn}(z_1,z_2;\vec{x}_1-\vec{x}_2)\,(z_2)^2\,{\cal V}_{0mn}(z_2,\vec{x}_2).
\label{hache1ijsimplied}
\end{equation}

Let us point out  that the simplification that it is achieved by doing the replacement in (\ref{varphi}) is a reflection of the fact that the part of the action in (\ref{classaction}) which depends on $\phi(x)$ is invariant under  the Weyl transformations
\begin{equation*}
g_{\mu\nu}\rightarrow(\Omega(x))^{-2} g_{\mu\nu},\quad \phi(x)\rightarrow \Omega(x)\,\phi(x).
\end{equation*}

\section{Putting the classical action on-shell.}

The AdS/CFT correspondence prescription --put forward in \cite{Witten:1998qj}-- to obtain the correlation functions of the conformal field theory at lowest order out of the partition function of the bulk theory, demands that one puts the classical action on shell. By putting the classical action on shell, one means that each field in that action is to be replaced with a solution to the classical equation of motion of the bulk theory: the solution must satisfy appropriate boundary conditions. In our case, we have to substitute $\phi^{(0)}(z,\vec{x})=z \varphi(z,\vec{x})$, $h^{(1)}_{ij}(z,\vec{x})$ and $\phi^{(1)}(z,\vec{x})$ --which are given in (\ref{varphi}), (\ref{hache1ijsimplied}) and (\ref{phione}), respectively-- in $S_{\text{\tiny{class}}}$, given in (\ref{classaction}), and keep only the contributions involving just four $\varphi^{(0)}(z,\vec{x})$.

We shall begin by computing the contributions that are obtained by setting $\lambda=0$ in $S_{\text{\tiny{class}}}$, i.e., the contributions that have a purely gravitational origin. These contributions are order $\kappa^2$ and to obtain them we shall work out first the contribution, say $S^{(2)}_{\text{HE}}[h^{(1)}_{ij}]$, coming from $S_{\text{HE}}$ which is quadratic in $h^{(1)}_{ij}(x)$. After some integration by parts, one gets
\begin{equation}
S^{(2)}_{\text{HE}}[h^{(1)}_{ij};a]=\cfrac{1}{2}\int_{a}^{\infty}dz \idthreex\,\sqrt{\bar{g}}\,h^{(1)ij}({\cal K}h^{(1)})_{ij}\,+\,{\cal B}_{\text{HE}}[h^{(1)}_{ij};a],
\label{S2HE}
\end{equation}
where $({\cal K}h^{(1)})_{ij}$ is given in (\ref{Koperator}) of appendix A and ${\cal B}_{\text{HE}}$ is the following remnant boundary term:
\begin{equation}
\begin{array}{l}
{{\cal B}_{\text{HE}}[h^{(1)}_{ij};a]=\cfrac{1}{2}\int_{a}^{\infty}dz \idthreex\,\sqrt{\bar{g}}\,[\bar{\nabla}_\mu(\bar{h}^{(1)}\bar{\nabla}^\mu\bar{h}^{(1)})+
\bar{\nabla}_\mu(h^{(1)\mu\nu}\bar{\nabla}_{\rho}h^{(1)\rho}_\nu)+\bar{\nabla}_\mu(h^{(1)\rho\nu}\bar{\nabla}_{\nu}h^{(1)\mu}_\rho)+}\\[8pt]
{\phantom{{\cal B}_{\text{HE}}[h^{(1)}_{ij};a]=\cfrac{1}{2}\int_{a}^{\infty}dz \idthreex\,\sqrt{\bar{g}}\,[}
\bar{\nabla}_\mu(h^{(1)\rho\sigma}\bar{\nabla}^{\mu}h^{(1)}_{\rho\sigma})+\bar{\nabla}_\mu(\bar{h}^{(1)}\bar{\nabla}_\nu h^{(1)\nu\mu})].}
\label{BHE}
\end{array}
\end{equation}
Note that $\bar{h}^{(1)}=\bar{g}^{\mu\nu}h^{(1)}_{\mu\nu}=\bar{g}^{ij}h^{(1)}_{ij}$. The reader should  bear in mind that $h^{(1)}_{z\mu}(z,\vec{x})=0$, for we are working in the axial gauge --see (\ref{axgauge}).

Next, we shall compute the contribution, say $S^{(1)}_{\text{\tiny{scalar}}}[h^{(1)}_{ij},\phi^{(0)};a]$, which arises from $S_{\text{\tiny{scalar}}}$ in (\ref{classaction}) and that is linear in $h^{(1)}_{ij}$. This contribution runs thus
\begin{equation}
S^{(1)}_{\text{\tiny{scalar}}}[h^{(1)}_{ij},\phi^{(0)};a]=-\kappa\int_{a}^{\infty}dz \idthreex\,\sqrt{\bar{g}\,}h^{(1)}_{ij}{\cal T}_{0}^{ij}+{\cal B}_{\text{\tiny{scalar}}} [h^{(1)}_{ij},\phi^{(0)};a],
\label{S1scalar}
\end{equation}
where
\begin{equation}
\begin{array}{l}
{{\cal B}_{\text{\tiny{scalar}}} [h^{(1)}_{ij},\phi^{(0)};a]=\kappa\int_{a}^{\infty}dz \idthreex\,\sqrt{\bar{g}}\,[-\cfrac{1}{8}\,\bar{\nabla}_\mu(\bar{h}^{(1)}\phi^{(0)}\bar{\nabla}^\mu\phi^{(0)})+
\cfrac{1}{4}\,\bar{\nabla}_\mu(h^{(1)\mu}_\nu\phi^{(0)}\bar{\nabla}^\nu\phi^{(0)})+}\\[8pt]
{\phantom{{\cal B}_{\text{\tiny{scalar}}} [h^{(1)}_{ij}\,\,}
\cfrac{1}{12}\,\bar{\nabla}_\mu((\phi^{(0)})^2\bar{\nabla}_\nu h^{(1)\nu\mu})-
\cfrac{1}{12}\,\bar{\nabla}_\mu(h^{(1)\mu\nu}\bar{\nabla}_{\nu}((\phi^{(0)})^2)-\cfrac{1}{12}\,\bar{\nabla}_\mu((\phi^{(0)})^2\bar{\nabla}^\mu\bar{h}^{(1)})+}\\[8pt]
{\phantom{{\cal B}_{\text{\tiny{scalar}}} [h^{(1)}_{ij}\,\,}\cfrac{1}{12}\,\bar{\nabla}_\mu(\bar{h}^{(1)}\bar{\nabla}^\mu(\phi^{(0)})^2)].}
\label{Bscalar}
\end{array}
\end{equation}

Finally, $S_{\text{GHY}}+ S_{\text{B}}$ in (\ref{classaction}) yields the following contributions quadratic in $h^{(1)}_{ij}(a,\vec{x})$:
\begin{equation}
\begin{array}{l}
{S^{(2)}_{\text{GHY+B}}[h^{(1)}_{ij};a]=\cfrac{1}{a^3}\,\idthreex\,\Big[ 2\,\big(-\cfrac{1}{4}\, h^{(1)i}_j(a,\vec{x})\,h^{(1)j}_i(a,\vec{x})+\cfrac{1}{8}\,
\bar{h}^{(1)}(a,\vec{x})\bar{h}^{(1)}(a,\vec{x})\big)}\\[8pt]
{\phantom{S^{(2)}_{\text{GHY}+\text{B}}[h^{(1)}_{ij};a]=\cfrac{2}{a^3}\,\idthreex\,\Big[ }
a\, h^{(1)i}_j(a,\vec{x})\partial_z h^{(1)j}_i(z,\vec{x})
-\cfrac{a}{2}\,\bar{h}^{(1)}(a,\vec{x})\partial_z\bar{h}^{(1)}(z,\vec{x})\Big]|_{z=a},}
\label{BGHY}
\end{array}
\end{equation}
where $h^{(1)i}_j=\bar{g}^{im}h^{(1)}_{mj}$.

We shall show in the appendix C that
\begin{equation*}
\lim_{a\rightarrow 0}{\cal B}_{\text{HE}}[h^{(1)}_{ij};a]=0,\,\lim_{a\rightarrow 0}{\cal B}_{\text{\tiny{scalar}}} [h^{(1)}_{ij},\phi^{(0)};a]=0,\,
\lim_{a\rightarrow 0}S^{(2)}_{\text{GHY+B}}[h^{(1)}_{ij};a]=0;
\end{equation*}
so we will drop ${\cal B}_{\text{HE}}$, ${\cal B}_{\text{\tiny{scalar}}}$ and $ S^{(2)}_{\text{GHY+B}}$ from now on.

Let us go back to  $S^{(2)}_{\text{HE}}[h^{(1)}_{ij};a]$, $S^{(1)}_{\text{\tiny{scalar}}}[h^{(1)}_{ij},\phi^{(0)};a]$ in (\ref{S2HE}) and (\ref{S1scalar}).
Then, we define
\begin{equation}
S^{(\text{4pt})}[a]=\cfrac{1}{2}\int_{a}^{\infty}dz \idthreex\,\sqrt{\bar{g}}\,h^{(1)ij}({\cal K}h^{(1)})_{ij}-\kappa\int_{a}^{\infty}dz \idthreex\,\sqrt{\bar{g}\,}h^{(1)}_{ij}{\cal T}_{0}^{ij},
\label{S4ptprev}
\end{equation}
for this is the only contribution of order $\kappa^2$, involving only four $\phi^{(0)}$ and coming from $S_{\text{\tiny{class}}}$  that yields a non vanishing result in the limit $a\rightarrow 0$. Of course, by computing the functional derivatives of  $S^{(\text{4pt})}[a]$ with regard to the boundary field $\phi_0({\vec{k}})$ in (\ref{phizero}) -see also (\ref{varphi}), one will obtain the tree-level correlation function for the scalar dual operator ${\cal O}(\vec{k})$. But before doing this, we have do an obvious simplification of definition of $S^{(\text{4pt})}[a]$ and, then,  carry out the integrals involved in its definition.

Taking into account (\ref{eqh1ij}), one concludes that $S^{(\text{4pt})}[a]$ in (\ref{S4ptprev}) is given by
\begin{equation}
S^{(\text{4pt})}[a]=-\cfrac{1}{2}\,\kappa\int_{a}^{\infty}dz \idthreex\,\sqrt{\bar{g}}\,h^{(1)}_{ij}{\cal T}_{0}^{ij}.
\label{S4ptsimpler}
\end{equation}

Lastly, it is not difficult to show the interaction $\lambda\phi^4$ interaction term in $S_{\text{\tiny{class}}}$ in (\ref{classaction}) adds to  $S^{(\text{4pt})}[a]$ the following contribution
\begin{equation}
-3\,\cfrac{\lambda}{4!}\,\int_{a}^{\infty}dz \idthreex\,\sqrt{\bar{g}}\,(\phi^{(0)}(z,\vec{x}))^4,
\label{lphi4cont}
\end{equation}
modulo terms which vanish as $a\rightarrow 0$.

\section{The computation of $S^{(\text{4pt})}[a]$ and the four-point correlation function.}

Let us recall that  $h^{(1)}_{ij}$ is given in (\ref{hache1ijsimplied}) and that (\ref{TtoV}) holds. Then, $S^{(\text{4pt})}$ in (\ref{S4ptsimpler}) can be recast into the form
\begin{equation}
S^{(\text{4pt})}[a]=-\cfrac{1}{2}\,\kappa^2\int\prod_{\alpha=1}^{4}\idthreeka\;(2\pi)^3\,\delta(\sum_{\alpha=1}^4 \vec{k}_\alpha)\; {\cal S}(\vec{k}_1,\vec{k}_2,\vec{k}_3,\vec{k}_4;a)\,
\phi_0(\vec{k}_1)\phi_0(\vec{k}_2)\phi_0(\vec{k}_3)\phi_0(\vec{k}_4),
\label{S4pt}
\end{equation}
where
\begin{equation}
{\cal S}(\vec{k}_1,\vec{k}_2,\vec{k}_3,\vec{k}_4;a)=
\int_{a}^{\infty}dz_1\int_{a}^{\infty}dz_2\;{\cal V}_{0ij}(z_1;\vec{k}_1,\vec{k}_2)\,(z_1)^2\, G_{ij,mn}(z_1,z_2;\vec{k_1}+\vec{k}_2)\,(z_2)^2\,
{\cal V}_{0mn}(z_2;\vec{k}_3,\vec{k}_4)
\label{calS}
\end{equation}
and
\begin{equation}
\begin{array}{l}
{{\cal V}_{0mn}(z;\vec{p},\vec{q})=-
\cfrac{1}{12}\,\delta_{mn}\,\partial_z f(z,\vec{p})\partial_z f(z,\vec{q})+}\\[8pt]
{
[\cfrac{1}{12}\,\delta_{mn}\,\vec{p}\cdot\vec{q}-
\,\cfrac{1}{6}\,(p_m q_n+p_n q_m)+\cfrac{1}{12}\,(p_m p_n+q_m q_n)]f(z,\vec{p})f(z,\vec{q}),}\\[12pt]
{f(z,\vec{p})=\cfrac{\sqrt{\frac{2}{\pi}}(k z)^{1/2}  K_{1/2}[k z]}{\sqrt{\frac{2}{\pi}}\,(k a)^{1/2}  K_{1/2}[k a]}.}
\label{Vzeromn}
\end{array}
\end{equation}

The computation of the right hand side of (\ref{calS}) has been carried out by using Form \cite{Ruijl:2017dtg} and Mathematica \cite{mathematica}. Actually, $a$ in (\ref{calS}) can be set to zero before doing any integration, for no divergences arise by doing so. We have obtained the following result for ${\cal S}(\vec{k}_1,\vec{k}_2,\vec{k}_3,\vec{k}_4;a=0)$
\begin{equation}
\begin{array}{l}
{{\cal S}(\vec{k}_1,\vec{k}_2,\vec{k}_3,\vec{k}_4)={\cal S}(\vec{k}_1,\vec{k}_2,\vec{k}_3,\vec{k}_4;a=0)}\\[8pt]
{\cfrac{1}{576 E^3 s^2}\Big\{\cfrac{(k_{12}^2 + 3 k_{12} k_{34} + k_{34}^2) (3 (k_{1} - k_{2})^2 -
     s) s (-3 (k_{3} - k_{4})^2 + s)}{k_{12} k_{34}}\, - }\\[8pt]
{
\cfrac{ 9}{(k_{12} k_{34} + (k_{12} + k_{34}) \sqrt{s} + s)^2}\,
  (k_{12} k_{34} + 2 E \sqrt{s} +
     s) \Big[k_{2}^4 (k_{3}^4 - 2\, k_{3}^2 (k_{4}^2 - 3\, s) + (k_{4}^2 - s)^2) +}\\[8pt]

{k_{1}^4 (k_{3}^4 + k_{4}^4 + 6\, k_{4}^2 s + s^2 - 2\, k_{3}^2 (k_{4}^2 + s)) +
     s^2 (k_{3}^4 + (k_{4}^2 - s)^2 + k_{3}^2 (6 k_{4}^2 - 2 s - 8 t) +}\\[8pt]
 {       8\, (-k_{4}^2 + s) t + 8\, t^2) -
     2\, k_{2}^2 s (-3 k_{3}^4 + (k_{4}^2 - s) (k_{4}^2 - s - 4\, t) +
        2\, k_{3}^2 (k_{4}^2 + s + 2\, t)) -}\\[8pt]
    { 2\, k_{1}^2 (k_{2}^2 ((k_{3}^2 - k_{4}^2)^2 + 2\, (k_{3}^2 + k_{4}^2) s - 3\, s^2) +
        s (k_{3}^4 - 3 k_{4}^4 + 2\, k_{4}^2 s + s^2 + 2\, k_{3}^2 (k_{4}^2 - s - 2 t) +}\\[8pt]
      {     4\, (k_{4}^2 + s) t))\Big] + \cfrac{1}{
  k_{12} k_{34}}\Big[2 k_{1} k_{2} s^2 (-3 (k_{3} - k_{4})^2 + s) +
    s^2 (-3\, (k_{3}^2 - k_{4}^2)^2 + k_{34}^2 s) +}\\[8pt]
{    3\, k_{2}^4 (3\, (k_{3}^2 - k_{4}^2)^2 + 6\, (3 k_{3}^2 - k_{4}^2) s - s^2) +
    3\, k_{1}^4 (3 k_{3}^4 + 3\, k_{4}^4 + 18 k_{4}^2 s - s^2 - 6\, k_{3}^2 (k_{4}^2 + s)) +}\\[8pt]
{    k_{2}^2\, s\, (54\, k_{3}^4 - 18\, k_{4}^4 - 6\, k_{3} k_{4} s + s^2 +
       9\, k_{4}^2 (3\, s + 8\, t) - 9\, k_{3}^2 (4\, k_{4}^2 + 5\, s + 8\, t)) +}\\[8pt]
 {   k_{1}^2 \Big(6\, k_{2}^2 (-3\, (k_{3}^2 - k_{4}^2)^2 - 6\, (k_{3}^2 + k_{4}^2) s + s^2) +
       s [-18\, k_{3}^4 + 54\, k_{4}^4 - 6\, k_{3} k_{4} s + s^2 +}\\[8pt]
  {       9\, k_{3}^2 (-4\, k_{4}^2 + 3\, s + 8\, t) - 9\, k_{4}^2\, (5\, s + 8\, t)]\Big)\Big]\Big\}.
}
\label{calSresult}
\end{array}
\end{equation}
Details of the computations leading to (\ref{calSresult}) can be found in appendix B. In the previous equation, we have used the following notation $k_i=\sqrt{\vec{k}_i\cdot\vec{k}_i}$, $i=1,2,3,4$,  $E=k_1+k_2+k_3+k_4$, $k_{12}=k_1+k_2$ and $k_{34}=k_3+k_4$; and we have imposed $\vec{k}_4=-\vec{k}_1-\vec{k}_2-\vec{k}_3$. The Mandelstam variables $s$, $t$ and $u$ are defined thus
$s=(\vec{k}_1+\vec{k}_2)^2$, $t=(\vec{k}_1+\vec{k}_3)^2$ and $u=(\vec{k}_1+\vec{k}_4)^2$.

Let us recall that $\cal{O}$ denotes the dual operator, in the AdS/CFT correspondence framework, of the bulk scalar field $\phi$ in (\ref{classaction}). Then,  the tree-level connected contribution to the four-point correlation function \cite{Ammon:2015wua} of the operator $\cal{O}$, when $\lambda=0$, is given by:
\begin{equation}
\begin{array}{l}
{\langle{\cal{O}}(\vec{k}_1){\cal{O}}(\vec{k}_2){\cal{O}}(\vec{k}_3){\cal{O}}(\vec{k}_4)\rangle^{(c)}[\lambda=0]=-\cfrac{\delta^4 S^{(\text{4pt})}[a=0]}{\delta\phi_0(\vec{k}_1)\delta\phi_0(\vec{k}_2)\delta\phi_0(\vec{k}_3)\delta\phi_0(\vec{k}_4)}\Bigg|_{\phi_0=0}=}\\[16pt]
{\phantom{\langle{\cal{O}}(\vec{k}_1){\cal{O}}(\vec{k}_2){\cal{O}}(\vec{k}_3}
4\kappa^{2}
(2\pi)^3\,\delta(\sum_{\alpha=1}^4 \vec{k}_\alpha)\,{\cal W}(\vec{k}_1,\vec{k}_2,\vec{k}_3,\vec{k}_4),}\\[12pt]
{{\cal W}(\vec{k}_1,\vec{k}_2,\vec{k}_3,\vec{k}_4)=
{\cal W}_t(\vec{k}_1,\vec{k}_2,\vec{k}_3,\vec{k}_4)+{\cal W}_t(\vec{k}_1,\vec{k}_2,\vec{k}_3,\vec{k}_4)+{\cal W}_u(\vec{k}_1,\vec{k}_2,\vec{k}_3,\vec{k}_4),}\\[8pt]
{{\cal W}_s(\vec{k}_1,\vec{k}_2,\vec{k}_3,\vec{k}_4)={\cal S}(\vec{k}_1,\vec{k}_2,\vec{k}_3,\vec{k}_4),\,{\cal W}_t(\vec{k}_1,\vec{k}_2,\vec{k}_3,\vec{k}_4)={\cal S}(\vec{k}_1,\vec{k}_3,\vec{k}_2,\vec{k}_4),}\\[8pt]
{{\cal W}_u(\vec{k}_1,\vec{k}_2,\vec{k}_3,\vec{k}_4)={\cal S}(\vec{k}_1,\vec{k}_4,\vec{k}_2,\vec{k}_3),}
\label{calW}
\end{array}
\end{equation}
where $S^{(\text{4pt})}[a]$ in given in (\ref{S4pt}) and ${\cal S}(\vec{k}_1,\vec{k}_2,\vec{k}_3,\vec{k}_4)$ is to be found in (\ref{calSresult}). The subscripts $s$, $t$ and $u$ in $W_s$, $W_t$ and $W_s$ refer to the fact that the latter can be interpreted  as coming from the exchange of a graviton along the $s$, $t$ and $u$ channels, respectively --see (\ref{calS}). Recall that $\lambda$ is the coupling constant of the scalar self-interaction and that $\lambda=0$ in (\ref{calW})

To obtain the result in (\ref{calW}), one has to take into account that ${\cal S}(\vec{k}_1,\vec{k}_2,\vec{k}_3,\vec{k}_4)$ is invariant under the following three independent exchanges:
\begin{equation*}
\vec{k}_1\longleftrightarrow \vec{k}_2,\quad\vec{k}_3\longleftrightarrow \vec{k}_4\quad\text{and}\quad(\vec{k}_1,\vec{k}_2)\longleftrightarrow(\vec{k}_3,\vec{k}_4).
\end{equation*}
Notice that these three invariances of ${\cal S}(\vec{k}_1,\vec{k}_2,\vec{k}_3,\vec{k}_4)$ is apparent when the latter is expressed as in (\ref{calS}), not so if we look at the
value of ${\cal S}(\vec{k}_1,\vec{k}_2,\vec{k}_3,\vec{k}_4)$ in (\ref{calSresult}). Indeed, to obtain (\ref{calSresult}), we have used $\vec{k}_4=-\vec{k}_1-\vec{k}_2-\vec{k}_3$.
And yet, we have checked the invariance under the exchanges above by running the appropriate Mathematica code.

Finally, from (\ref{lphi4cont}), one can derive the contribution that must be added to\\ $\langle{\cal{O}}(\vec{k}_1){\cal{O}}(\vec{k}_2){\cal{O}}(\vec{k}_3){\cal{O}}(\vec{k}_4)\rangle^{(c)}[\lambda=0]$ in (\ref{calW}) to obtain the complete tree-level contribution to the four-point function of the dual operator ${\cal O}(\vec{k})$. The contribution in question reads
\begin{equation}
3\,\lambda(2\pi)^3\,\delta(\sum_{\alpha=1}^4 \vec{k}_\alpha)\;\cfrac{1}{E},
\label{lambdaboundary}
\end{equation}
where $E=k_1+k_2+k_3+k_4$.

\section{Conformal Ward identities.}

In this section, we shall check that ${\cal W}(\vec{k}_1,\vec{k}_2,\vec{k}_3,\vec{k}_4)$ in (\ref{calW}) satisfies the conformal Ward identities in momentum space obtained in \cite{Bzowski:2013sza}. Recall that $\vec{k}_4=-\vec{k}_1-\vec{k}_2-\vec{k}_3$ in ${\cal W}$. We should  notice that it has been shown in \cite{Oh:2020izq} that the contribution in (\ref{lambdaboundary}) satisfies the conformal Ward identities, a result that we have verified ourselves.

Next, let us begin with the conformal Ward identity implied by invariance under dilatations. According to \cite{Bzowski:2013sza}, this identity  reads
\begin{equation}
\sum_{\alpha=1}^3\,k_{\alpha}^i\,\cfrac{\partial{\cal W}(\vec{k}_1,\vec{k}_2,\vec{k}_3,\vec{k}_4)}{\partial k_\alpha^i}\,=\,(4\,\Delta-3\, d)\,{\cal W}(\vec{k}_1,\vec{k}_2,\vec{k}_3,\vec{k}_4),
\label{dilatationwi}
\end{equation}
where $\Delta$ is the conformal dimension of ${\cal O}(\vec{x})$ and $d$ is the dimension of the boundary space. Now, just by power counting --see (\ref{calSresult}), one concludes that
\begin{equation*}
{\cal W}(M\vec{k}_1,M\vec{k}_2,M\vec{k}_3,M\vec{k}_4)=\cfrac{1}{M}\,{\cal W}(\vec{k}_1,\vec{k}_2,\vec{k}_3,\vec{k}_4),
\end{equation*}
which obviously implies (\ref{dilatationwi}), for in our case $\Delta=2$ and $d=3$.

Let us move on and consider the Ward identity implying invariance under special conformal transformations. But, first, let us introduce the following objects
\begin{equation*}
\begin{array}{l}
{b^j C_1^j=b^j\sum_{\alpha=1}^3\,k_\alpha^j\,\cfrac{\partial^2}{\partial k_{\alpha}^i\partial k_{\alpha}^i}{\cal W}(\vec{k}_1,\vec{k}_2,\vec{k}_3,\vec{k}_4),}\\[8pt]
{b^j C_2^j=b^j\sum_{\alpha=1}^3\,k_\alpha^i\,\cfrac{\partial^2}{\partial k_{\alpha}^i\partial k_{\alpha}^j}{\cal W}(\vec{k}_1,\vec{k}_2,\vec{k}_3,\vec{k}_4),}\\[8pt]
{b^j C_3^j=b^j\sum_{\alpha=1}^3\,(-2)\,\cfrac{\partial}{\partial k_\alpha^j}{\cal W}(\vec{k}_1,\vec{k}_2,\vec{k}_3,\vec{k}_4).}
\end{array}
\end{equation*}
Then, the special conformal transformation Ward identity reads \cite{Bzowski:2013sza}:
\begin{equation}
b^j\,(C_1^j-2\,C_2^j\,+\,C_3^j)=0,
\label{theCsequation}
\end{equation}
where $b^j$ is the infinitesimal vector defining the infinitesimal especial conformal transformations.

We have not been able to verify that (\ref{theCsequation}) holds for arbitrary $\vec{k}_1$, $\vec{k}_2$ and $\vec{k}_3$ by using Mathematica, let alone by hand, because huge algebraic expressions demanded to be simplified. What we have done, by asking Mathematica to carry out the simplifications for us within a sensible span of time --shorter than an hour in a standard Intel Core i7 laptop, is to check that (\ref{theCsequation}) is indeed satisfied for the following sets of momenta:
\begin{equation*}
\begin{array}{l}
{\{\vec{k}_1=(a,0,0),\,\vec{k}_2=(b,0,0),\,\vec{k}_3=(c,0,0)\},\,\{\vec{k}_1=(a,0,0),\,\vec{k}_2=(b,0,0),\,\vec{k}_3=(0,c,0)\},}\\[8pt]
{\{\vec{k}_1=(a,0,0),\,\vec{k}_2=(b,0,0),\,\vec{k}_3=(0,0,c)\},\,\{\vec{k}_1=(a,0,0),\,\vec{k}_2=(0,b,0),\,\vec{k}_3=(0,c,0)\},}\\[8pt]
{\{\vec{k}_1=(a,0,0),\,\vec{k}_2=(0,b,0),\,\vec{k}_3=(0,0,c)\},\,\{\vec{k}_1=(a,0,0),\,\vec{k}_2=(0,0,b),\,\vec{k}_3=(0,0,c)\}.}
\end{array}
\end{equation*}
$a,\,b$ and $c$ are arbitrary real numbers. We believe that to display any partial result leading to (\ref{theCsequation}) is quite useless, since, e.g., $C_1^1$ involves   more than three hundred terms: $a$, $b$ and $c$ being arbitrary. Actually, what we have done with help from Mathematica is to show that (\ref{theCsequation}) holds for each ${\cal W}_s$, ${\cal W}_t$ and ${\cal W}_u$ in (\ref{calW}), individually. Let us give just one instance of such a cancellation:  replace $W$ with $W_u$ in (\ref{theCsequation}), then, you get
\begin{equation*}
\begin{array}{l}
{C_1^1\,=\,\cfrac{55003 - \frac{658845}{\sqrt{182}} - \sqrt{\frac{ 291533218261}{182} - 802824594 \sqrt{\frac{26}{7}}}}{4567992},}\\[8pt]
{C_2^1\,=\,\cfrac{10140760 + 2635380 \sqrt{182} -
 3\,\sqrt{14\, (8846269483913 + 590279440080 \sqrt{182})}}{3325498176},}\\[8pt]
{C_3^1\,=\,\cfrac{-898212 - 670824 \sqrt{13} - 406403 \sqrt{14} +
 359370 \sqrt{182}}{151159008},}\\[8pt]
{ C_1^1-2 C_2^1+ C_3^1=0.}
\end{array}
\end{equation*}
if $\vec{k}_1=(1,0,0)$, $\vec{k}_2=(0,2,0)$ and $\vec{k}_3=(0,0,3)$.

We have also checked the numerical consistency with zero of (\ref{theCsequation}) in the following sense. We have used the random real number generator of Mathematica to give values to the components of $\vec{k}_1$, $\vec{k}_2$ and $\vec{k}_3$. The random number generator has chosen real numbers between $-100$ and $100$ with a precision of 20 digits. We have then worked out the left hand side of (\ref{theCsequation}) and so obtained a result which consistent with zero --i.e. $0\pm 10^{-21}$--. We have repeated the process 10 times,
always reaching the same conclusion. Let us just give an instance of these computations. If
\begin{equation*}
\begin{array}{l}
{\vec{k}_1=(2.1927121967338693989,81.454072629674493406,-85.047391618404869870),}\\[4pt]
{\vec{k}_2=(-98.092197908364545822,89.788465437796389444,-76.905960631110550703),}\\[4pt]
{\vec{k}_3=(34.020467731854881945,-23.860053648929931141,48.477525138035992674),}
\end{array}
\end{equation*}
then,
\begin{equation*}
\begin{array}{l}
{C_1^1=12774334731901\times 10^{-20},\,C_2^1=3272604694993\times 10^{-20},}\\[4pt]
{C_3^1=-6229125341915\times 10^{-20}.}
\end{array}
\end{equation*}
The previous values lead to $C_1^1-2\, C_2^1+C_3^1=0\pm 10^{-21}$.

\section{Closing comments.}
The main results of this paper are in section 4, namely, in (\ref{calSresult}), (\ref{calW}) and (\ref{lambdaboundary}). These results yield the tree-level four-point boundary correlation function for a scalar field conformally coupled to Einstein's gravity on Euclidean $\text{AdS}_4$. We checked that our results are consistent -see section 5-- with conformal invariance, as befits the AdS/CFT correspondence. Further, the content of this paper confirms the usefulness of the momentum space techniques to obtain explicit expression for correlation functions in Euclidean AdS. Last but not least, we believe that our results will be of  help to study non-gaussian corrections to primordial fluctuations generated during inflation.

\section{Acknowledgements.}
I heartily thank J. Anero and E. \'Alvarez for illuminating discussions. This piece research work has been  financially supported  in part by the Spanish Ministry of Science, Innovation and Universities through grant PID2023-149834NB-I00.

\newpage

\appendix
\section{First order corrections.}

Let us first recall that we are working in the axial gauge for graviton field: $h_{z\mu}(x)=0$, $\forall \mu=z,i$, $i=1,2,3$. Let ${\cal K}$ denote a differential operator  whose action on $h_{ij}(x)$ is the following
\begin{equation}
\begin{array}{l}
{({\cal K}h)_{ij}=\bar{\nabla}_\mu\bar{\nabla}^\mu h_{ij}-\bar{g}_{ij} \bar{\nabla}_\mu\bar{\nabla}^\mu(\bar{g}^{\rho\sigma}h_{\rho\sigma})+
\bar{g}_{ij}\bar{\nabla}_\mu\bar{\nabla}_\nu h^{\mu\nu}+\cfrac{1}{2}\,(\bar{\nabla}_i\bar{\nabla}_j+ \bar{\nabla}_j\bar{\nabla}_i)(\bar{g}^{\rho\sigma}h_{\rho\sigma})-}\\[4pt]
{\phantom{({\cal K}h)_{ij}=\,}
\bar{\nabla}_i\bar{\nabla}_\nu h^{\nu}_j-\bar{\nabla}_j\bar{\nabla}_\nu h^{\nu}_i+ 2\,h_{ij}+\bar{g}_{ij}\,\bar{g}^{mn}h_{mn}.}
\label{Koperator}
\end{array}
\end{equation}

With an eye on solving the equations of motion,  which come from the action in (\ref{classaction}), by doing a formal double series expansion in $\kappa$ and $\lambda$, we shall write those equations of motion as follows:
\begin{equation}
\begin{array}{l}
{({\cal K}h)_{ij}=\kappa\,{\cal T}_{ij}+\kappa\, S^{(3)}_{ij}(h)+\,\text{higher-order terms in $\lambda$ and $\kappa$},}\\[8pt]
{-\bar{\nabla}_\mu\bar{\nabla}^\mu\phi-2\phi=-\cfrac{\lambda}{3!}(\phi)^3+\,\text{higher-order terms in $\lambda$ and $\kappa$},}
\label{eom}
\end{array}
\end{equation}
where $S^{(3)}_{ij}(h)$ contains three $h^{(1)}_{ij}$'s and no $\phi$, and
\begin{equation*}
{\cal T}_{ij}=\cfrac{1}{4}\,\bar{g}_{ij}\,\phi^2+\cfrac{1}{6}\,\bar{g}_{ij}\phi \bar{\nabla}_\mu\bar{\nabla}^\mu\phi-
\cfrac{1}{12}\,\bar{g}_{ij}\bar{\nabla}_\mu\phi \bar{\nabla}^\mu\phi+\cfrac{1}{3}\,\bar{\nabla}_i\phi \bar{\nabla}_j\phi
-\cfrac{1}{12}\,\phi(\bar{\nabla}_i\bar{\nabla}_j+\bar{\nabla}_j\bar{\nabla}_i)\phi.
\end{equation*}

Let $h^{(0)}_{ij}(x)$ and $\phi^{(0)}(x)$ be the solution to the equations of motion in (\ref{eom}) when $\kappa$ and $\lambda$ are both set to zero. Then, in view of (\ref{eom}), there is a perturbative correction of order $\kappa$, say $h^{(1)}_{ij}(x)$, to $h^{(0)}_{ij}(x)$, which is obtained by solving the equation:
\begin{equation}
({\cal K}h^{(1)})_{ij}=\kappa{\cal T}_{0ij}.
\label{eqh1ij}
\end{equation}
In the previous equation, the action of ${\cal K}$ on $h^{(1)}_{ij}(x)$ is obtained by replacing $h_{ij}$ with $h^{(1)}_{ij}$ in $(\ref{Koperator})$ and ${\cal T}_{0ij}$ is given by
\begin{equation*}
\begin{array}{l}
{{\cal T}_{0ij}=\cfrac{1}{4}\,\bar{g}_{ij}\,(\phi^{(0)})^2+\cfrac{1}{6}\,\bar{g}_{ij}\phi^{(0)}\bar{\nabla}_\mu\bar{\nabla}^\mu\phi^{(0)}-
\cfrac{1}{12}\,\bar{g}_{ij}\bar{\nabla}_\mu\phi^{(0)}\bar{\nabla}^\mu\phi^{(0)}+\cfrac{1}{3}\,\bar{\nabla}_i\phi^{(0)}\bar{\nabla}_j\phi^{(0)}
-}\\[4pt]
{\phantom{{\cal T}_{0ij}=}\,\cfrac{1}{12}\,\phi^{(0)}(\bar{\nabla}_j\bar{\nabla}_i+\bar{\nabla}_j\bar{\nabla}_i)\phi^{(0)}.}
\end{array}
\end{equation*}

Let $G_{ij,mn}(z_1,z_2;\vec{x}_1-\vec{x}_2)$ be the Green function of the operator ${\cal K}$ in (\ref{Koperator}):
\begin{equation*}
({\cal K}G)_{ij,mn}(x_1,x_2)=\cfrac{1}{2}\,(\bar{g}_{im}(x_1)\bar{g}_{jn}(x_1)+\bar{g}_{in}(x_1)\bar{g}_{jm}(x_1))
\cfrac{1}{\sqrt{\bar{g}(x_1)}}\,\delta(z_1-z_2)\,\delta(\vec{x}_1-\vec{x}_2),
\end{equation*}
where $x_1=(z_1,\vec{x}_1)$ and $x_2=(z_2,\vec{x}_2)$. Then
\begin{equation*}
h^{(1)}_{ij}(z_1,x_1)=\kappa\int_{a}^{\infty}dz_2\idxtwo\,\sqrt{\bar{g}(z_2,\vec{x}_2)}\,G_{ij,mn}(z_1,z_2;\vec{x}_1-\vec{x}_2)\,{\cal T}_0^{mn}(z_2,\vec{x}_2)
\end{equation*}
solves (\ref{eqh1ij}). $G_{ij,mn}(z_1,z_2;\vec{x}_1-\vec{x}_2)$ is, by definition,  the graviton bulk-to-bulk propagator and it is  given in (\ref{BtBpropGR}).

We shall introduce now the Green function, $G(z_1,z_2;\vec{x}_1-\vec{x}_2)$, of the differential operator $-(\bar{\nabla}_\mu\bar{\nabla}^\mu+2)$:
\begin{equation*}
-(\bar{\nabla}_\mu\bar{\nabla}^\mu+2)G(z_1,z_2;\vec{x}_1-\vec{x}_2)=\cfrac{1}{\sqrt{\bar{g}(x_1)}}\,\delta(z_1-z_2)\,\delta(\vec{x}_1-\vec{x}_2).
\end{equation*}
$G(z_1,z_2;\vec{x}_1-\vec{x}_2)$ is, of course, the bulk-to-bulk propagator of the scalar field $\phi(x)$ and  its Fourier transform can be found in (\ref{phiprops}).
Then
\begin{equation*}
\phi^{(1)}(z,\vec{x})=-\cfrac{\lambda}{3!}\,\int_{a}^{\infty}dz_2\,\idxtwo\,\sqrt{\bar{g}(z_2,\vec{x}_2)}\, G(z_1,z_2;\vec{x}_1-\vec{x}_2)\,(\phi^{(0)}(z_2,\vec{x}_2))^3,
\end{equation*}

is a perturbative correction to $\phi^{(0)}$ of order $\lambda$, for it solves
\begin{equation*}
-\bar{\nabla}_\mu\bar{\nabla}^\mu\phi^{(1)}-2\phi^{(1)}=-\cfrac{\lambda}{3!}(\phi^{(0)})^3.
\end{equation*}

\section{Details of the computations leading to ${\cal S}(\vec{k}_1,\vec{k}_2,\vec{k}_3,\vec{k}_4)$.}

Let $G^{(i)}_{i_1 j_1,i_2 j_2}(z_1,z_2;\vec{k})$, $i=1,2,3$, be defined as follows
\begin{equation*}
G^{(i)}_{i_1 j_1,i_2 j_2}(z_1,z_2;\vec{k})=(z_1 z_ 2)^{-1/2}\int_{0}^{\infty} d\omega\,\omega\, J_{3/2}[\omega z_1]\,{\tilde G}^{(i)}_{i_1 j_1,i_2 j_2}(\omega,\vec{k})\,J_{3/2}[\omega z_2],
\end{equation*}
where
\begin{equation*}
\begin{array}{l}
{{\tilde G}^{(1)}_{i_1 j_1,i_2 j_2}(\omega,\vec{k})=G_1(T_{i_1 i_2} T_{j_1 j_2}+T_{i_1 j_2} T_{j_1 i_2}-T_{i_1 j_1} T_{i_2 j_2}),}\\[8pt]
{{\tilde G}^{(2)}_{i_1 j_1,i_2 j_2}(\omega,\vec{k})=G_2(T_{i_1 i_2} L_{j_1 j_2}+L_{i_1 i_2} T_{j_1 j_2}+T_{i_1 j_2} L_{j_1 i_2}+L_{i_1 j_2} T_{j_1 i_2}  -T_{i_1 j_1} L_{i_2 j_2}-L_{i_1 j_1} T_{i_2 j_2}+}\\[4pt]
{\phantom{{\tilde G}^{(2)}_{i_1 j_1,i_2 j_2}(\omega,\vec{k})=}
L_{i_1 i_2} L_{j_1 j_2}  +L_{i_1 j_2} L_{j_1 i_2}-L_{i_1 j_1} L_{i_2 j_2}),}\\[4pt]
{{\tilde G}^{(2)}_{i_1 j_1,i_2 j_2}(\omega,\vec{k})= G_3(L_{i_1 i_2} L_{j_1 j_2}+L_{i_1 j_2} L_{j_1 i_2}-L_{i_1 j_1} L_{i_2 j_2})
}
\end{array}
\end{equation*}
and  $G_i$, $i=1,2,3$ and $T_{ij}$ and $L_{ij}$ are given in (\ref{coefBtBpropGR}). Then, by taking into account (\ref{BtBpropGR}) and (\ref{calS}), one concludes that
\begin{equation}
{\cal S}(\vec{k}_1,\vec{k}_2,\vec{k}_3,\vec{k}_4)=\text{VG}^{(1)}\text{V}+\text{VG}^{(2)}\text{V}+\text{VG}^{(3)}\text{V},
\label{sumofVGVs}
\end{equation}
where
\begin{equation*}
\text{VG}^{(i)}\text{V}=\int_{0}^{\infty}\!\!\!dz_1\int_{0}^{\infty}\!\!\!dz_2\;{\cal V}_{0ij}(z_1;\vec{k}_1,\vec{k}_2)\,(z_1)^2\, G^{(i)}_{ij,mn}(z_1,z_2;\vec{k_1}+\vec{k}_2)\,(z_2)^2\,
{\cal V}_{0mn}(z_2;\vec{k}_3,\vec{k}_4),\, i\!=\!1,2,3.
\end{equation*}

Next, let us introduce the following integrals
\begin{equation*}
\begin{array}{l}
{\text{FLFRG}^{(i)}=\int_{0}^\infty\,d\omega \omega {\tilde G}^{(i)}(\omega,\vec{k}_1+\vec{k}_2)\times}\\[4pt]
{\phantom{\text{FLFRG}^{(i)}=}
\int_0^\infty dz_1 \sqrt{\frac{2}{\pi}} \sqrt{k_1 z_1}\, K_{1/2}[  k_1 z_1] \sqrt{\frac{2}{\pi}} \sqrt{k_2 z_1} K_{1/2}[  k_2 z_1] z_ 1^{3/2} J_{3/2}[ w z_1]\times}\\[4pt]
{\phantom{\text{FLFRG}^{(i)}=}
\int_0^\infty dz_2 \sqrt{\frac{2}{\pi}} \sqrt{k_3 z_2}\, K_{1/2}[  k_3 z_2] \sqrt{\frac{2}{\pi}} \sqrt{k_4 z_2} K_{1/2}[  k_4 z_2] z_ 2^{3/2} J_{3/2}[ w z_2],}\\[8pt]
{\text{DFLFRG}^{(i)}=\int_{0}^\infty\,d\omega\, \omega\, {\tilde G}^{(i)}(\omega,\vec{k}_1+\vec{k}_2)\times}\\[4pt]
{\phantom{\text{DFLFRG}^{(i)}=}\int_0^\infty dz_1 [\partial_{z_1}\{\sqrt{\frac{2}{\pi}} \sqrt{k_1 z_1}\, K_{1/2}[  k_1 z_1]\}][\partial_{z_1}\{\sqrt{\frac{2}{\pi}}\sqrt{k_2 z_1} K_{1/2}[  k_2 z_1]\}]
z_ 1^{3/2} J_{3/2}[ w z_1]\times}\\[4pt]
{\phantom{\text{DFLFRG}^{(i)}=}
\int_0^\infty dz_2 \sqrt{\frac{2}{\pi}} \sqrt{k_3 z_2}\, K_{1/2}[  k_3 z_2] \sqrt{\frac{2}{\pi}}\sqrt{k_4 z_2} K_{1/2}[  k_4 z_2] z_ 2^{3/2} J_{3/2}[ w z_2],}\\[8pt]
{\text{FLDFRG}^{(i)}=\int_{0}^\infty\,d\omega \omega {\tilde G}^{(i)}(\omega,\vec{k}_1+\vec{k}_2)\times}\\[4pt]
{\phantom{\text{FLDFRG}^{(i)}=}
\int_0^\infty dz_1 \sqrt{\frac{2}{\pi}} \sqrt{k_1 z_1}\, K_{1/2}[  k_1 z_1] \sqrt{\frac{2}{\pi}} \sqrt{k_2 z_1} K_{1/2}[  k_2 z_1] z_ 1^{3/2} J_{3/2}[ w z_1]\times}\\[4pt]
{\phantom{\text{FLDFRG}^{(i)}=}
\int_0^\infty dz_2 [\partial_{z_2}\{\sqrt{\frac{2}{\pi}} \sqrt{k_3 z_2}\, K_{1/2}[  k_3 z_2]\}][\partial_{z_2}\{ \sqrt{\frac{2}{\pi}} \sqrt{k_4 z_2} K_{1/2}[  k_4 z_2]\}]
 z_ 2^{3/2} J_{3/2}[ w z_2],}\\[8pt]
{\text{DFLDFRG}^{(i)}=\int_{0}^\infty\,d\omega \omega {\tilde G}^{(i)}(\omega,\vec{k}_1+\vec{k}_2)\times}\\[4pt]
{\phantom{\text{DFLDFRG}^{(i)}=}
\int_0^\infty dz_1[\partial_{z_1}\{ \sqrt{\frac{2}{\pi}} \sqrt{k_1 z_1}\, K_{1/2}[  k_1 z_1]\}][\partial_{z_1}\{ \sqrt{\frac{2}{\pi}} \sqrt{k_2 z_1} K_{1/2}[  k_2 z_1]\}]
 z_ 1^{3/2} J_{3/2}[ w z_1]\times}\\[4pt]
{\phantom{\text{DFLDFRG}^{(i)}=}
\int_0^\infty dz_2 [\partial_{z_2}\{\sqrt{\frac{2}{\pi}} \sqrt{k_3 z_2}\, K_{1/2}[  k_3 z_2]\}][\partial_{z_2}\{ \sqrt{\frac{2}{\pi}} \sqrt{k_4 z_2} K_{1/2}[  k_4 z_2]\}]
 z_ 2^{3/2} J_{3/2}[ w z_2],}
\end{array}
\end{equation*}
where $i=1,2,3$. We have worked out the previous integrals with help from Mathematica
\begin{equation}
\begin{array}{l}
{\text{FLFRG}^{(1)} = -\cfrac{(k_1 + k_2) (k_3 + k_4) + 2\, (k_1 + k_2 + k_3 + k_4) \sqrt{s} +
   s}{(k_1 + k_2 + k_3 + k_4)^3 (k_1 + k_2 + \sqrt{s})^2 (k_3 + k_4 + \sqrt{
     s})^2 s^2},}\\[8pt]
{\text{FLFRG}^{(2)} = -\cfrac{1}{(k_1 + k_2) (k_3 + k_4) (k_1 + k_2 + k_3 + k_4)^3 s^2},}\\[8pt]
{\text{FLFRG}^{(3)} = -\cfrac{
  k_1^2 + 2\, k_1 k_2 + k_2^2 + 3\, k_1 (k_3 + k_4) +
   3\, k_2 (k_3 + k_4) + (k_3 + k_4)^2}{(k_1 + k_2)^3 (k_3 + k_4)^3 (k_1 + k_2 +
     k_3 + k_4)^3 s},}\\[8pt]
{\text{FLDFRG}^{(2)} = -\cfrac{k_3 k_4}{(k_1 + k_2) (k_3 + k_4) (k_1 + k_2 + k_3 + k_4)^3 s^2},}\\[8pt]
{\text{FLDFRG}^{(3)} = -\cfrac{
  k_3 k_4 (k_1^2 + 2\, k_1 k_2 + k_2^2 + 3\, k_1 (k_3 + k_4) +
     3\, k_2 (k_3 + k_4) + (k_3 + k_4)^2)}{(k_1 + k_2)^3 (k_3 + k_4)^3 (k_1 +
     k_2 + k_3 + k_4)^3 s},}\\[8pt]
{\text{DFLFRG}^{(2)} = -\cfrac{k_1 k_2}{(k_1 + k_2) (k_3 + k_4) (k_1 + k_2 + k_3 + k_4)^3 s^2},}\\[8pt]
{\text{DFLFRG}^{(3)} = -\cfrac{
  k_1 k_2 (k_1^2 + 2\, k_1 k_2 + k_2^2 + 3\, k_1 (k_3 + k_4) +
     3\, k_2 (k_3 + k_4) + (k_3 + k_4)^2)}{(k_1 + k_2)^3 (k_3 + k_4)^3 (k_1 +
     k_2 + k_3 + k_4)^3 s},}\\[8pt]
{\text{DFLDFRG}^{(2)} = -\cfrac{
  k_1 k_2 k_3 k_4}{(k_1 + k_2) (k_3 + k_4) (k_1 + k_2 + k_3 + k_4)^3 s^2},}\\[8pt]
{\text{DFLDFRG}^{(3)} = -\cfrac{k_1 k_2 k_3 k_4 (k_1^2 + 2\, k_1 k_2 + k_2^2 + 3\, k_1 (k_3 + k_4) +
        3\, k_2 (k_3 + k_4) + (k_3 + k_4)^2)}{(k_1 + k_2)^3 (k_3 + k_4)^3 (k_1 +
        k_2 + k_3 + k_4)^3 s},}
\label{integralvalues}
\end{array}
\end{equation}
where $k_\alpha=\sqrt{\vec{k}_\alpha^2}$, $\alpha=1,2,3,4$. Notice that we have not listed neither the value of $\text{DFLFRG}^{(1)}$ nor the value of $\text{FLDFRG}^{(1)}$
nor the value of $\text{DFLDFRG}^{(1)}$, since, as we shall see below, they do not contribute to $\text{VG}^{(1)}\text{V}$ due to fact that
\begin{equation*}
(T_{i_1 i_2} T_{j_1 j_2}+T_{i_1 j_2} T_{j_1 i_2}-T_{i_1 j_1} T_{i_2 j_2})\delta_{i_1 j_1}=0.
\end{equation*}

We have carried out the tensor contractions involved in the computation of $\text{VG}^{(i)}\text{V}$, $i=1,2,3$ by using FORM \cite{Ruijl:2017dtg}. The results that we have obtained, upon imposing $\vec{k}_4=-\sum_{i=1}^3\,\vec{k}_i$, are given next:
\begin{equation}
\begin{array}{l}
{\text{VG}^{(1)}\text{V}=}\\[4pt]
{\text{FLFRG}^{(1)}\, (\frac{1}{64} s^2 u^2 - \frac{3}{32} s^2 t u + \frac{1}{64} s^2 t^2 +
    \frac{1}{16} k_3^2 s^2 u + \frac{1}{16} k_3^2 s^2 t + \frac{1}{16} k_3^2 s^3 -
    \frac{1}{16} k_3^4 s^2 - \frac{1}{32} k_2^2 s u^2 +}\\[4pt]
{\frac{1}{16} k_2^2 s t u + \frac{3}{32} k_2^2 s t^2 - \frac{1}{32} k_2^2 s^2 u + \frac{3}{32} k_2^2 s^2 t -
    \frac{1}{4} k_2^2 k_3^2 s t - 3/16 k_2^2 k_3^2 s^2 + \frac{1}{8} k_2^2 k_3^4 s +
    \frac{1}{64} k_2^4 u^2 +}\\[4pt]
{    \frac{1}{32} k_2^4 t u + \frac{1}{64} k_2^4 t^2 + \frac{1}{16} k_2^4 s u -
    \frac{1}{16} k_2^4 s t + \frac{1}{64} k_2^4 s^2 - \frac{1}{16} k_2^4 k_3^2 u -
    \frac{1}{16} k_2^4 k_3^2 t + \frac{1}{16} k_2^4 k_3^2 s + \frac{1}{16} k_2^4 k_3^4 -}\\[4pt]
{    \frac{1}{32} k_2^6 u - \frac{1}{32} k_2^6 t -
    \frac{1}{32} k_2^6 s + \frac{1}{16} k_2^6 k_3^2 +
    \frac{1}{64} k_2^8 + \frac{3}{32} k_1^2 s u^2 + \frac{1}{16} k_1^2 s t u - \frac{1}{32} k_1^2 s t^2 +
    \frac{3}{32} k_1^2 s^2 u - \frac{1}{32} k_1^2 s^2 t - }\\[4pt]
{     \frac{1}{4} k_1^2 k_3^2 s u - 3/16 k_1^2 k_3^2 s^2 + \frac{1}{8} k_1^2 k_3^4 s - \frac{1}{32} k_1^2 k_2^2 u^2 -
    \frac{1}{16} k_1^2 k_2^2 t u - \frac{1}{32} k_1^2 k_2^2 t^2 - \frac{1}{4} k_1^2 k_2^2 s u -
    \frac{1}{4} k_1^2 k_2^2 s t - }\\[4pt]
{    \frac{3}{32} k_1^2 k_2^2 s^2 + \frac{1}{8} k_1^2 k_2^2 k_3^2 u +
    \frac{1}{8} k_1^2 k_2^2 k_3^2 t + \frac{3}{8} k_1^2 k_2^2 k_3^2 s -
    \frac{1}{8} k_1^2 k_2^2 k_3^4 + \frac{1}{32} k_1^2 k_2^4 u + \frac{1}{32} k_1^2 k_2^4 t +
    \frac{5}{32} k_1^2 k_2^4 s - }\\[4pt]
    {\frac{1}{16} k_1^2 k_2^4 k_3^2 + \frac{1}{64} k_1^4 u^2 +
    \frac{1}{32} k_1^4 t u + \frac{1}{64} k_1^4 t^2 - \frac{1}{16} k_1^4 s u + \frac{1}{16} k_1^4 s t +
    \frac{1}{64} k_1^4 s^2 - \frac{1}{16} k_1^4 k_3^2 u -   \frac{1}{16} k_1^4 k_3^2 t +}\\[4pt]
{    \frac{1}{16} k_1^4 k_3^2 s + \frac{1}{16} k_1^4 k_3^4 + \frac{1}{32} k_1^4 k_2^2 u +
    \frac{1}{32} k_1^4 k_2^2 t + \frac{5}{32} k_1^4 k_2^2 s - \frac{1}{16} k_1^4 k_2^2 k_3^2 -
    \frac{1}{32} k_1^4 k_2^4 - \frac{1}{32} k_1^6 u - \frac{1}{32} k_1^6 t -}\\[4pt]
{     \frac{1}{32} k_1^6 s + \frac{1}{16} k_1^6 k_3^2 + \frac{1}{64} k_1^8),}
\label{VG1V}
\end{array}
\end{equation}
\begin{equation}
\begin{array}{l}
{\text{VG}^{(2)}\text{V}=}\\[4pt]
{ \text{DFLDFRG}^{(2)}\, (-\frac{1}{48} s^2)
  +\text{ DFLFRG}^{(2)}\, (\frac{1}{96} s^2 u + \frac{1}{96} s^2 t + \frac{1}{144} s^3 - \frac{1}{96} k_2^2 s^2 -
     \frac{1}{96} k_1^2 s^2)+}\\[4pt]
{ \text{FLDFRG}^{(2)}\,(-\frac{1}{288} s^3 + \frac{1}{96} k_2^2 s^2 + \frac{1}{96} k_1^2 s^2) +}\\[4pt]
 {\text{FLFRG}^{(2)}\, (\frac{1}{192} s^2 u^2 + \frac{1}{96} s^2 t u + \frac{1}{192} s^2 t^2 +
     \frac{5}{576} s^3 u + \frac{5}{576} s^3 t + \frac{1}{288} s^4 - \frac{1}{48} k_3^2 s^2 u -
     \frac{1}{48} k_3^2 s^2 t - }\\[4pt]
{     \frac{1}{48} k_3^2 s^3 + \frac{1}{48} k_3^4 s^2 +
     \frac{1}{32} k_2^2 s u^2 - \frac{1}{16} k_2^2 s t u - \frac{3}{32} k_2^2 s t^2 +
     \frac{1}{192} k_2^2 s^2 u - \frac{23}{192} k_2^2 s^2 t - \frac{5}{192} k_2^2 s^3 +}\\[4pt]
{     \frac{1}{4} k_2^2 k_3^2 s t +   \frac{7}{48} k_2^2 k_3^2 s^2 - \frac{1}{8} k_2^2 k_3^4 s -
     \frac{1}{64} k_2^4 u^2 - \frac{1}{32} k_2^4 t u - \frac{1}{64} k_2^4 t^2 - \frac{1}{16} k_2^4 s u +
     \frac{1}{16} k_2^4 s t + \frac{1}{96} k_2^4 s^2 +}\\[4pt]
{ \frac{1}{16} k_2^4 k_3^2 u +
     \frac{1}{16} k_2^4 k_3^2 t - \frac{1}{16} k_2^4 k_3^2 s - \frac{1}{16} k_2^4 k_3^4 +
     \frac{1}{32} k_2^6 u + \frac{1}{32} k_2^6 t + \frac{1}{32} k_2^6 s - \frac{1}{16} k_2^6 k_3^2 -
     \frac{1}{64} k_2^8 - \frac{3}{32} k_1^2 s u^2 -}\\[4pt]
{      \frac{1}{16} k_1^2 s t u +
     \frac{1}{32} k_1^2 s t^2 - \frac{23}{192} k_1^2 s^2 u + \frac{1}{192} k_1^2 s^2 t -
     \frac{5}{192} k_1^2 s^3 + \frac{1}{4} k_1^2 k_3^2 s u + \frac{7}{48} k_1^2 k_3^2 s^2 -
     \frac{1}{8} k_1^2 k_3^4 s +}\\[4pt]
{     \frac{1}{32} k_1^2 k_2^2 u^2 + \frac{1}{16} k_1^2 k_2^2 t u +
     \frac{1}{32} k_1^2 k_2^2 t^2 + \frac{1}{4} k_1^2 k_2^2 s u + \frac{1}{4} k_1^2 k_2^2 s t +
     \frac{3}{16} k_1^2 k_2^2 s^2 - \frac{1}{8} k_1^2 k_2^2 k_3^2 u -
     \frac{1}{8} k_1^2 k_2^2 k_3^2 t - }\\[4pt]
{     \frac{3}{8} k_1^2 k_2^2 k_3^2 s +
     \frac{1}{8} k_1^2 k_2^2 k_3^4 - \frac{1}{32} k_1^2 k_2^4 u - \frac{1}{32} k_1^2 k_2^4 t -
     \frac{5}{32} k_1^2 k_2^4 s + \frac{1}{16} k_1^2 k_2^4 k_3^2 - \frac{1}{64} k_1^4 u^2 -
     \frac{1}{32} k_1^4 t u - }\\[4pt]
{     \frac{1}{64} k_1^4 t^2 + \frac{1}{16} k_1^4 s u - \frac{1}{16} k_1^4 s t +
     \frac{1}{96} k_1^4 s^2 + \frac{1}{16} k_1^4 k_3^2 u + \frac{1}{16} k_1^4 k_3^2 t -
     \frac{1}{16} k_1^4 k_3^2 s - \frac{1}{16} k_1^4 k_3^4 - \frac{1}{32} k_1^4 k_2^2 u -}\\[4pt]
{     \frac{1}{32} k_1^4 k_2^2 t - \frac{5}{32} k_1^4 k_2^2 s + \frac{1}{16} k_1^4 k_2^2 k_3^2 +
     \frac{1}{32} k_1^4 k_2^4 + \frac{1}{32} k_1^6 u + \frac{1}{32} k_1^6 t + \frac{1}{32} k_1^6 s -
     \frac{1}{16} k_1^6 k_3^2 - \frac{1}{64} k_1^8),}
\label{VG2V}
\end{array}
\end{equation}
and
\begin{equation}
\begin{array}{l}
{\text{VG}^{(3)}\text{V}=}\\[4pt]
{\text{DFLDFRG}^{(3)}\, (\frac{1}{144} s^2)
   + \text{DFLFRG}^{(3)}\, (-\frac{1}{96} s u^2 - \frac{1}{48} s t u - \frac{1}{96} s t^2 - \frac{5}{288} s^2 u -
      \frac{5}{288} s^2 t - \frac{1}{144} s^3 + }\\[4pt]
{      \frac{1}{24} k_3^2 s u +
      \frac{1}{24} k_3^2 s t + \frac{1}{24} k_3^2 s^2 - \frac{1}{24} k_3^4 s + \frac{1}{48} k_2^2 s u + \frac{1}{48} k_2^2 s t +
      \frac{5}{288} k_2^2 s^2 - \frac{1}{24} k_2^2 k_3^2 s - \frac{1}{96} k_2^4 s +
      \frac{1}{48} k_1^2 s u + }\\[4pt]
{      \frac{1}{48} k_1^2 s t + \frac{5}{288} k_1^2 s^2 -
      \frac{1}{24} k_1^2 k_3^2 s - \frac{1}{48} k_1^2 k_2^2 s - \frac{1}{96} k_1^4 s)
   +\text{ FLDFRG}^{(3)}\, (\frac{1}{288} k_2^2 s^2 - \frac{1}{96} k_2^4 s + \frac{1}{288} k_1^2 s^2 +}\\[4pt]
{      \frac{1}{48} k_1^2 k_2^2 s -
      \frac{1}{96} k_1^4 s)
   +\text{ FLFRG}^{(3)}\, (-\frac{1}{192} k_2^2 s u^2 - \frac{1}{96} k_2^2 s t u - \frac{1}{192} k_2^2 s t^2 -
      \frac{5}{576} k_2^2 s^2 u - \frac{5}{576} k_2^2 s^2 t - }\\[4pt]
{      \frac{1}{288} k_2^2 s^3 +
      \frac{1}{48} k_2^2 k_3^2 s u + \frac{1}{48} k_2^2 k_3^2 s t + \frac{1}{48} k_2^2 k_3^2 s^2 -
      \frac{1}{48} k_2^2 k_3^4 s + \frac{1}{64} k_2^4 u^2 + \frac{1}{32} k_2^4 t u +
      \frac{1}{64} k_2^4 t^2 + \frac{7}{192} k_2^4 s u + }\\[4pt]
{      \frac{7}{192} k_2^4 s t +
      \frac{11}{576} k_2^4 s^2 - \frac{1}{16} k_2^4 k_3^2 u - \frac{1}{16} k_2^4 k_3^2 t -
      \frac{1}{12} k_2^4 k_3^2 s + \frac{1}{16} k_2^4 k_3^4 - \frac{1}{32} k_2^6 u - \frac{1}{32} k_2^6 t -
      \frac{1}{32} k_2^6 s + }\\[4pt]
{      \frac{1}{16} k_2^6 k_3^2 + \frac{1}{64} k_2^8 - \frac{1}{192} k_1^2 s u^2 -
      \frac{1}{96} k_1^2 s t u - \frac{1}{192} k_1^2 s t^2 - \frac{5}{576} k_1^2 s^2 u -
      \frac{5}{576} k_1^2 s^2 t - \frac{1}{288} k_1^2 s^3 +}\\[4pt]
{       \frac{1}{48} k_1^2 k_3^2 s u +
      \frac{1}{48} k_1^2 k_3^2 s t + \frac{1}{48} k_1^2 k_3^2 s^2 - \frac{1}{48} k_1^2 k_3^4 s -
      \frac{1}{32} k_1^2 k_2^2 u^2 - \frac{1}{16} k_1^2 k_2^2 t u - \frac{1}{32} k_1^2 k_2^2 t^2 -
      \frac{1}{32} k_1^2 k_2^2 s u -}\\[4pt]
{ \frac{1}{32} k_1^2 k_2^2 s t - \frac{1}{288} k_1^2 k_2^2 s^2 +
      \frac{1}{8} k_1^2 k_2^2 k_3^2 u + \frac{1}{8} k_1^2 k_2^2 k_3^2 t +
      \frac{1}{12} k_1^2 k_2^2 k_3^2 s - \frac{1}{8} k_1^2 k_2^2 k_3^4 + \frac{1}{32} k_1^2 k_2^4 u +
      \frac{1}{32} k_1^2 k_2^4 t +}\\[4pt]
{       \frac{1}{96} k_1^2 k_2^4 s - \frac{1}{16} k_1^2 k_2^4 k_3^2 +
      \frac{1}{64} k_1^4 u^2 + \frac{1}{32} k_1^4 t u + \frac{1}{64} k_1^4 t^2 +
      \frac{7}{192} k_1^4 s u + \frac{7}{192} k_1^4 s t + \frac{11}{576} k_1^4 s^2 -
   \frac{1}{16} k_1^4 k_3^2 u -}\\[4pt]
{ \frac{1}{16} k_1^4 k_3^2 t - \frac{1}{12} k_1^4 k_3^2 s +
      \frac{1}{16} k_1^4 k_3^4 + \frac{1}{32} k_1^4 k_2^2 u + \frac{1}{32} k_1^4 k_2^2 t +
      \frac{1}{96} k_1^4 k_2^2 s - \frac{1}{16} k_1^4 k_2^2 k_3^2 - \frac{1}{32} k_1^4 k_2^4 -
      \frac{1}{32} k_1^6 u - }\\[4pt]
{ \frac{1}{32} k_1^6 t - \frac{1}{32} k_1^6 s + \frac{1}{16} k_1^6 k_3^2 +
      \frac{1}{64} k_1^8).}
\label{VG3V}
\end{array}
\end{equation}

The substitution of (\ref{integralvalues}) in (\ref{VG1V}), (\ref{VG2V}) and (\ref{VG3V}) leads to (\ref{calSresult}), once (\ref{sumofVGVs}) is used. Of course, all these computations could not have been feasible without using Mathematica.

\section{The vanishing of the boundary contributions.}

From (\ref{hache1ijsimplied}), one derives the following result
\begin{equation*}
\begin{array}{l}
{h^{(1)}_{ij}(a,\vec{x})=\kappa\,\idthreek\idfourk\,e^{i\vec{x}(\vec{k}_3+\vec{k}_4)}\,h^{(1)}_{ij}(a;\vec{k}_3,\vec{k}_4),}\\[4pt]
{h^{(1)}_{ij}(a;\vec{k}_3,\vec{k}_4)=
\int_a^\infty dz_ 2\,G_{ij,mn}(a,z_2;\vec{k}_3+\vec{k}_4)\,(z_2)^2\,
{\cal V}_{0mn}(z_2;\vec{k}_3,\vec{k}_4)\phi_{0}(\vec{k}_3)\phi_{0}(\vec{k}_4).}
\end{array}
\end{equation*}
${\cal V}_{0mn}$ is defined in (\ref{Vzeromn}) and $G_{ij,mn}$ is given in (\ref{BtBpropGR}).

By explicit computation of the integrals and, then, by doing a series expansion around $a=0$, one shows that
\begin{equation*}
h^{(1)}_{ij}[a,\vec{k}_3,\vec{k}_4]\,=\,a\, t_{ij}(\vec{k}_3,\vec{k}_4)\phi_{0}(\vec{k}_3)\phi_{0}(\vec{k}_4)+ O(a^2),
\end{equation*}
where $t_{ij}(\vec{k}_3,\vec{k}_4)$ is a tensor which is diverges linearly as $k_3$ or $k_4$ go to infinity and $\phi_{0}(\vec{k}_3)$ and $\phi_{0}(\vec{k}_4)$ go to zero at infinity fast enough. Hence,
\begin{equation*}
h^{(1)}_{ij}[a,\vec{x}]=a\,f_{ij}(\vec{x})+O(a^2).
\end{equation*}

The previous result, also leads to
\begin{equation}
\begin{array}{l}
{\bar{\nabla}_z h^{(1)}_{ij}[z,\vec{x}]|_{z=a}=\big[\partial_z +\frac{2}{z}\,]h^{(1)}_{ij}[z,\vec{x}]|_{z=a}=3\,f_{ij}(\vec{x})+O(a),}\\[8pt]
{\bar{\nabla}_i h^{(1)}_{jz}[z,\vec{x}]|_{z=a}=\cfrac{1}{a}\,h^{(1)}_{ij}[a,\vec{x}]=\,f_{ij}(\vec{x})+O(a).}
\label{asympder}
\end{array}
\end{equation}
To obtain the first result in the previous equations and to check the consistency of our computations, we have computed first $\partial_z h^{(1)}_{ij}[z,\vec{x}]$, then, set $z=a$ and, finally, we have series expanded around $a=0$.

Taking into account  (\ref{asympder}), that $\bar{g}(a,\vec{x})=1/a^8$ and  that $\bar{g}^{\mu\nu}(a,\vec{x})=a^2 \delta^{\mu\nu}$, one easily concludes that
\begin{equation}
\begin{array}{l}
{-\int_{a}^{\infty}dz \idthreex\,\sqrt{\bar{g}}\,\bar{\nabla}_\mu(\bar{h}^{(1)}\bar{\nabla}^\mu\bar{h}^{(1)})=\idthreex\sqrt{\bar{g}}\,(\bar{h}^{(1)}\bar{\nabla}^z\bar{h}^{(1)})|_{z=a}=
\idthreex\sqrt{\bar{g}}\,\bar{g}^{ij}\bar{g}^{zz}\bar{g}^{mn}(\bar{h}^{(1)}_{ij}\bar{\nabla}_z\bar{h}^{(1)}_{mn})|_{z=a}}\\[8pt]
{\phantom{-\int_{a}^{\infty}dz \idthreex\,\sqrt{\bar{g}}\,}
=a^{3}\,\text{constant}+O(a^4)\rightarrow 0\quad\text{as}\quad
 a\rightarrow 0.}
\label{firstzero}
\end{array}
\end{equation}
Analogously,
\begin{equation}
\begin{array}{l}
{-\int_{a}^{\infty}dz \idthreex\,\sqrt{\bar{g}}\,\bar{\nabla}_\mu(h^{(1)\rho\nu}\bar{\nabla}_{\nu}h^{(1)\mu}_\rho)=
 \idthreex\,\sqrt{\bar{g}}\,(h^{(1)\rho\nu}\bar{\nabla}_{\nu}h^{(1)z}_\rho)|_{z=a}=}\\[8pt]   {\idthreex\,\sqrt{\bar{g}}\,\bar{g}^{zz}\bar{g}^{ij}\bar{g}^{mn}(h^{(1)}_{im}\bar{\nabla}_{n}h^{(1)}_{jz})\vert_{z=a}=a^{3}\,\text{constant}+O(a^4)\rightarrow 0\quad\text{as}\quad
 a\rightarrow 0.}
 \label{secondzero}
 \end{array}
\end{equation}
Proceeding as in (\ref{firstzero}) and (\ref{secondzero}), one shows that the remaining contributions to ${\cal B}_{\text{HE}}[h^{(1)}_{ij};a]$ in (\ref{BHE}) also vanish as
$a^3$ when $a\rightarrow 0$. Thus, we conclude that
\begin{equation*}
\lim_{a\rightarrow 0}{\cal B}_{\text{HE}}[h^{(1)}_{ij};a]=0.
\end{equation*}

Next, let us show that
\begin{equation*}
\lim_{a\rightarrow 0}{\cal B}_{\text{\tiny{scalar}}} [h^{(1)}_{ij},\phi^{(0)};a]=0,
\end{equation*}
${\cal B}_{\text{\tiny{scalar}}}$ is displayed in (\ref{Bscalar}). Let us recall the $\phi^{(0)}(z,\vec{x})$ is given in (\ref{varphi}). Then, taking into account that  $\phi^{(0)}(a,\vec{x})=a\, \phi_0(\vec{x})$ and that
$\partial_z\phi^{(0)}(z,\vec{x})|_{z=a}=\phi_0(\vec{x})$, one concludes that
\begin{equation*}
\begin{array}{l}
{\int_{a}^{\infty}dz \idthreex\,\sqrt{\bar{g}}\bar{\nabla}_\mu(\bar{h}^{(1)}\phi^{(0)}\bar{\nabla}^\mu\phi^{(0)})=
-\idthreex\,\sqrt{\bar{g}}\bar{h}^{(1)}\phi^{(0)}\bar{\nabla}^z\phi^{(0)}|_{z=a}=}\\[8pt]
{-\idthreex\,\sqrt{\bar{g}}\bar{g}^{zz}\bar{g}^{ij}h^{(1)}_{ij}\phi^{(0)}\partial_z\phi^{(0)}|_{z=a}= a^2\,\text{constant}+O(a^3),}\\[12pt]
{\int_{a}^{\infty}dz \idthreex\,\sqrt{\bar{g}}\bar{\nabla}_\mu((\phi^{(0)})^2\bar{\nabla}_\nu h^{(1)\nu\mu})=-\idthreex\,\sqrt{\bar{g}}(\phi^{(0)})^2\bar{\nabla}_\nu h^{(1)\nu z}|_{z=a}=}\\[8pt]
{\idthreex\,\sqrt{\bar{g}}\bar{g}^{ij}\bar{g}^{zz}(\phi^{(0)})^2\bar{\nabla}_i h^{(1)}_{jz}|_{z=a}=-\idthreex\,\sqrt{\bar{g}}\bar{g}^{ij}\bar{g}^{zz}(\phi^{(0)})^2\cfrac{1}{z} h^{(1)}_{jz}|_{z=a}=}\\[8pt]
{ a^{-4} a^2 a^2 a^2 a^{-1} a\,\text{constant}+O(a^3)=a^2 \,\text{constant}+O(a^3).}
\end{array}
\end{equation*}
The same process leads to the conclusion that each of the remaining terms contributing to ${\cal B}_{\text{\tiny{scalar}}}$  in (\ref{Bscalar}) go to zero as a constant
times $a^2$.

Let us point out that the fact the contributions to ${\cal B}_{\text{HE}}[h^{(1)}_{ij};a]$ go to  zero as $c_{HE}\,a^3$ whereas the contributions to ${\cal B}_{\text{\tiny{scalar}}}$  in (\ref{Bscalar}) vanish as $c_{S}\,a^2$ is just a matter of dimensional constistency. Indeed, the quotient between the constants $c_{HE}$ and
$c_{S}$, $c_{HE}/c_{S}$, has the dimensions of a mass.

There remains to prove that
\begin{equation*}
\lim_{a\rightarrow 0}S^{(2)}_{\text{GHY+B}}[h^{(1)}_{ij};a]=0,
\end{equation*}
where $S^{(2)}_{\text{GHY+B}}[h^{(1)}_{ij};a]$ can be found in (\ref{BGHY}).
\begin{equation*}
\begin{array}{l}
{\cfrac{1}{a^3}\,\idthreex h^{(1)i}_j(a,\vec{x})\,h^{(1)j}_i(a,\vec{x})=\cfrac{1}{a^3}\,\idthreex \bar{g}^{ij}\bar{g}^{mn}h^{(1)}_{im}(a,\vec{x})\,h^{(1)}_{jn}(a,\vec{x})=}\\[8pt]
{a^{-3} a^2 a^2 a^2\,\text{constant}+O(a^4)=a^3\,\text{constant}+O(a^4),}\\[12pt]
{\cfrac{1}{a^3}\,\idthreex\, a\,\bar{g}^{ij} h^{(1)i}_{j}(a,\vec{x})\partial_z h^{(1)j}_i(a,\vec{x})1|_{z=a}=}\\[8pt]
{\cfrac{1}{a^3}\,\idthreex\, a\,\bar{g}^{ij} h^{(1)}_{im}(a,\vec{x})\partial_z(\bar{g}^{mn} h^{(1)}_{jn}(z,\vec{x}))|_{z=a}=}\\[8pt]
{a^{-3} a a^2 a a^2\,\text{constant}+O(a^4)=a^3\,\text{constant}+O(a^4),}
\end{array}
\end{equation*}
where we have used that
\begin{equation*}
\partial_z(\bar{g}^{mn} h^{(1)}_{jn}(z,\vec{x}))|_{z=a}=\delta^{mn}(2 a h^{(1)}_{jn}(a,\vec{x}))+ a^2\partial_z h^{(1)}_{jn}(z,\vec{x}))|_{z=a}=
\delta^{mn} a^{2}\,3\,f_{jn}(\vec{x})+O(a^3).
\end{equation*}


\newpage

\begin{thebibliography}{99}





\bibitem{Liu:1998ty}
H.~Liu and A.~A.~Tseytlin,
Phys. Rev. D \textbf{59} (1999), 086002
doi:10.1103/PhysRevD.59.086002
[arXiv:hep-th/9807097 [hep-th]].

\bibitem{Freedman:1998bj}
D.~Z.~Freedman, S.~D.~Mathur, A.~Matusis and L.~Rastelli,
Phys. Lett. B \textbf{452} (1999), 61-68
doi:10.1016/S0370-2693(99)00229-4
[arXiv:hep-th/9808006 [hep-th]].

\bibitem{DHoker:1999kzh}
E.~D'Hoker, D.~Z.~Freedman, S.~D.~Mathur, A.~Matusis and L.~Rastelli,
Nucl. Phys. B \textbf{562} (1999), 353-394
doi:10.1016/S0550-3213(99)00525-8
[arXiv:hep-th/9903196 [hep-th]].

\bibitem{Poland:2018epd}
D.~Poland, S.~Rychkov and A.~Vichi,
Rev. Mod. Phys. \textbf{91} (2019), 015002
doi:10.1103/RevModPhys.91.015002
[arXiv:1805.04405 [hep-th]].

\bibitem{Dolan:2003hv}
F.~A.~Dolan and H.~Osborn,
Nucl. Phys. B \textbf{678} (2004), 491-507
doi:10.1016/j.nuclphysb.2003.11.016
[arXiv:hep-th/0309180 [hep-th]].


\bibitem{Osborn:1993cr}
H.~Osborn and A.~C.~Petkou,
Annals Phys. \textbf{231}, 311-362 (1994)
doi:10.1006/aphy.1994.1045
[arXiv:hep-th/9307010 [hep-th]].

\bibitem{Bzowski:2013sza}
A.~Bzowski, P.~McFadden and K.~Skenderis,
JHEP \textbf{03} (2014), 111
doi:10.1007/JHEP03(2014)111
[arXiv:1304.7760 [hep-th]].


\bibitem{Bzowski:2015pba}
A.~Bzowski, P.~McFadden and K.~Skenderis,
JHEP \textbf{03} (2016), 066
doi:10.1007/JHEP03(2016)066
[arXiv:1510.08442 [hep-th]].

\bibitem{Coriano:2019sth}
C.~Corian\`o and M.~M.~Maglio,
JHEP \textbf{09} (2019), 107
doi:10.1007/JHEP09(2019)107
[arXiv:1903.05047 [hep-th]].

\bibitem{Bzowski:2019kwd}
A.~Bzowski, P.~McFadden and K.~Skenderis,
Phys. Rev. Lett. \textbf{124} (2020) no.13, 131602
doi:10.1103/PhysRevLett.124.131602
[arXiv:1910.10162 [hep-th]].


\bibitem{Coriano:2019nkw}
C.~Corian\`o, M.~M.~Maglio and D.~Theofilopoulos,
Eur. Phys. J. C \textbf{80} (2020) no.6, 540
doi:10.1140/epjc/s10052-020-8089-1
[arXiv:1912.01907 [hep-th]].


\bibitem{Bzowski:2020kfw}
A.~Bzowski, P.~McFadden and K.~Skenderis,
JHEP \textbf{01} (2021), 192
doi:10.1007/JHEP01(2021)192
[arXiv:2008.07543 [hep-th]].


\bibitem{Coriano:2020ees}
C.~Corian\`o and M.~M.~Maglio,
Phys. Rept. \textbf{952} (2022), 1-95
doi:10.1016/j.physrep.2021.11.005
[arXiv:2005.06873 [hep-th]].




\bibitem{Raju:2011mp}
S.~Raju,
Phys. Rev. D \textbf{83} (2011), 126002
doi:10.1103/PhysRevD.83.126002
[arXiv:1102.4724 [hep-th]].

\bibitem{Raju:2012zs}
S.~Raju,
Phys. Rev. D \textbf{85} (2012), 126008
doi:10.1103/PhysRevD.85.126008
[arXiv:1201.6452 [hep-th]].

\bibitem{Jatkar:2013uga}
D.~P.~Jatkar and J.~H.~Oh,
JHEP \textbf{10} (2013), 170
doi:10.1007/JHEP10(2013)170
[arXiv:1305.2008 [hep-th]].





\bibitem{Albayrak:2019asr}
S.~Albayrak, C.~Chowdhury and S.~Kharel,
JHEP \textbf{10} (2019), 274
doi:10.1007/JHEP10(2019)274
[arXiv:1904.10043 [hep-th]].



\bibitem{Albayrak:2019yve}
S.~Albayrak and S.~Kharel,
JHEP \textbf{12} (2019), 135
doi:10.1007/JHEP12(2019)135
[arXiv:1908.01835 [hep-th]].

\bibitem{Oh:2020zvm}
J.~H.~Oh,
JHEP \textbf{11} (2020), 100
doi:10.1007/JHEP11(2020)100
[arXiv:2005.08521 [hep-th]].

\bibitem{Albayrak:2020fyp}
S.~Albayrak, S.~Kharel and D.~Meltzer,
JHEP \textbf{03} (2021), 249
doi:10.1007/JHEP03(2021)249
[arXiv:2012.10460 [hep-th]].



\bibitem{Bzowski:2022rlz}
A.~Bzowski, P.~McFadden and K.~Skenderis,
JHEP \textbf{12} (2022), 039
doi:10.1007/JHEP12(2022)039
[arXiv:2207.02872 [hep-th]].

\bibitem{Armstrong:2023phb}
C.~Armstrong, H.~Goodhew, A.~Lipstein and J.~Mei,
JHEP \textbf{08} (2023), 206
doi:10.1007/JHEP08(2023)206
[arXiv:2304.07206 [hep-th]].

\bibitem{Chowdhury:2023khl}
C.~Chowdhury and K.~Singh,
JHEP \textbf{12} (2023), 109
doi:10.1007/JHEP12(2023)109
[arXiv:2305.18529 [hep-th]].


\bibitem{Bzowski:2023jwt}
A.~Bzowski,
JHEP \textbf{04} (2024), 082
doi:10.1007/JHEP04(2024)082
[arXiv:2312.11625 [hep-th]].



\bibitem{Albayrak:2023kfk}
S.~Albayrak, S.~Kharel and X.~Wang,
JHEP \textbf{07} (2024), 281
doi:10.1007/JHEP07(2024)281
[arXiv:2312.02154 [hep-th]].


\bibitem{Anero:2024bob}
J.~Anero and C.~P.~Martin,
[arXiv:2410.07365 [hep-th]].

\bibitem{Maldacena:2002vr}
J.~M.~Maldacena,
JHEP \textbf{05} (2003), 013
doi:10.1088/1126-6708/2003/05/013
[arXiv:astro-ph/0210603 [astro-ph]].

\bibitem{McFadden:2011kk}
P.~McFadden and K.~Skenderis,
JCAP \textbf{06} (2011), 030
doi:10.1088/1475-7516/2011/06/030
[arXiv:1104.3894 [hep-th]].


\bibitem{Ghosh:2014kba}
A.~Ghosh, N.~Kundu, S.~Raju and S.~P.~Trivedi,
JHEP \textbf{07} (2014), 011
doi:10.1007/JHEP07(2014)011
[arXiv:1401.1426 [hep-th]].



\bibitem{Armstrong:2023phb}
C.~Armstrong, H.~Goodhew, A.~Lipstein and J.~Mei,
JHEP \textbf{08} (2023), 206
doi:10.1007/JHEP08(2023)206
[arXiv:2304.07206 [hep-th]].

\bibitem{Chowdhury:2023arc}
C.~Chowdhury, A.~Lipstein, J.~Mei, I.~Sachs and P.~Vanhove,
[arXiv:2312.13803 [hep-th]].

\bibitem{deHaro:2006ymc}
S.~de Haro, I.~Papadimitriou and A.~C.~Petkou,
Phys. Rev. Lett. \textbf{98} (2007), 231601
doi:10.1103/PhysRevLett.98.231601
[arXiv:hep-th/0611315 [hep-th]].



\bibitem{Witten:1998qj}
E.~Witten,
Adv. Theor. Math. Phys. \textbf{2} (1998), 253-291
doi:10.4310/ATMP.1998.v2.n2.a2
[arXiv:hep-th/9802150 [hep-th]].


\bibitem{Liu:1998bu}
H.~Liu and A.~A.~Tseytlin,
Nucl. Phys. B \textbf{533} (1998), 88-108
doi:10.1016/S0550-3213(98)00443-X
[arXiv:hep-th/9804083 [hep-th]].


\bibitem{Albayrak:2020isk}
S.~Albayrak, C.~Chowdhury and S.~Kharel,
Phys. Rev. D \textbf{101} (2020) no.12, 124043
doi:10.1103/PhysRevD.101.124043
[arXiv:2001.06777 [hep-th]].

\bibitem{Albayrak:2023kfk}
S.~Albayrak, S.~Kharel and X.~Wang,
JHEP \textbf{07} (2024), 281
doi:10.1007/JHEP07(2024)281
[arXiv:2312.02154 [hep-th]].

\bibitem{Ammon:2015wua}
M.~Ammon and J.~Erdmenger,
Cambridge University Press, 2015,
SBN 978-1-107-01034-5, 978-1-316-23594-2




\bibitem{Ruijl:2017dtg}
B.~Ruijl, T.~Ueda and J.~Vermaseren,
``FORM version 4.2,''
[arXiv:1707.06453 [hep-ph]].


\bibitem{mathematica}
Wolfram Research, Inc, Mathematica 12.3, Champaign, IL (2021)

\bibitem{Oh:2020izq}
J.~H.~Oh,
[arXiv:2001.05379 [hep-th]].

\end{thebibliography}
\end{document}